\documentstyle [12pt] {article}

\newcommand{\gtwid}{\mathrel{\raise.3ex\hbox{$>$\kern-.75em\lower1ex
\hbox{$\sim$}}}}
\newcommand{\ltwid}{\mathrel{\raise.3ex\hbox{$<$\kern-.75em\lower1ex
\hbox{$\sim$}}}}
\newcommand{\beq}{\begin{equation}}
\newcommand{\eeq}{\end{equation}}
\newcommand{\beqs}{\begin{eqnarray}}
\newcommand{\eeqs}{\end{eqnarray}}

\catcode`@=11
\@addtoreset{equation}{section}
\@addtoreset{equation}{subsection}
\def\theequation{\ifnum\value{section}=0 \arabic{equation}\ignorespaces
\else \ifnum\value{section}=-1 A.\arabic{equation}\ignorespaces
\else \ifnum\value{subsection}=0 \thesection.\arabic{equation}\ignorespaces
\else \thesection.\arabic{subsection}.\arabic{equation}\ignorespaces
                           \fi
                      \fi
                 \fi}
\catcode`@=12

\begin{document}

\def\thefootnote{\fnsymbol{footnote}}
\baselineskip 6.0mm

\begin{flushright}
\begin{tabular}{l}
ITP-SB-95-23    \\
June, 1995
\end{tabular}
\end{flushright}

\vspace{8mm}
\begin{center}
{\Large \bf Complex-Temperature Properties of the 2D Ising Model for
Nonzero Magnetic Field}

\vspace{3mm}

\setcounter{footnote}{0}
Victor Matveev\footnote{email: vmatveev@max.physics.sunysb.edu}
\setcounter{footnote}{6}
and Robert Shrock\footnote{email: shrock@max.physics.sunysb.edu}

\vspace{6mm}
Institute for Theoretical Physics  \\
State University of New York       \\
Stony Brook, N. Y. 11794-3840  \\

\vspace{16mm}

{\bf Abstract}
\end{center}

    We study the complex-temperature phase diagram of the square-lattice
 Ising model for nonzero external magnetic field $H$, i.e. for $0 \le \mu \le
\infty$, where $\mu=e^{-2\beta H}$. We also carry out a similar analysis for
$-\infty \le \mu \le 0$.  The results for the interval $-1 \le \mu \le 1$
provide a new way of continuously connecting the two known exact solutions of
this model, viz., at $\mu=1$ (Onsager, Yang) and $\mu=-1$ (Lee and Yang).  Our
methods include calculations of complex-temperature zeros of the partition
function and analysis of low-temperature series expansions.  For real nonzero
$H$, the inner branch of a lima\c{c}on bounding the FM phase breaks and forms
two complex-conjugate arcs.  We study the singularities and associated
exponents of thermodynamic functions at the endpoints of these arcs.  For $\mu
< 0$, there are two line segments of singularities on the negative and positive
$u$ axis, and we carry out a similar study of the behavior at the inner
endpoints of these arcs, which constitute the nearest singularities to the
origin in this case.  Finally, we also determine the exact
complex-temperature phase diagrams at $\mu=-1$ on the honeycomb
and triangular lattices and discuss the relation between these and the
corresponding zero-field phase diagrams.

\vspace{16mm}

\pagestyle{empty}
\newpage

\pagestyle{plain}
\pagenumbering{arabic}
\renewcommand{\thefootnote}{\arabic{footnote}}
\setcounter{footnote}{0}

\section{Introduction}
\label{intro}

       The two-dimensional Ising model serves as a prototype of a
statistical mechanical system which undergoes a phase transition with
associated spontaneous symmetry breaking and long range order.
The free energy of the (spin $1/2$) Ising model was first
calculated by Onsager \cite{ons}, and the expression for the spontaneous
magnetization first derived by Yang \cite{yang} (both for the square
lattice).  However, the model has never been solved in an arbitrary nonzero
external magnetic field, and this has long remained an outstanding open
problem.  Hence, any additional information that one can gain about
the Ising model in a magnetic field is of value.
In elucidating the properties of the zero-field model, it has proved
useful to generalize the temperature variable to complex
values.  There are several reasons for this.  First, one can
understand more deeply the physical behavior of various
thermodynamic quantities by seeing how they vary as analytic functions of
complex temperature (CT).
Second, one can see how the physical phases of a given
model generalize to regions in appropriate complex-temperature variables.
Third, a knowledge of the complex-temperature singularities of quantities which
have not been calculated exactly helps in the search for exact, closed-form
expressions for these quantities.  For the (spin 1/2) zero-field Ising model on
the square lattice the complex-temperature zeros of the partition function
were first discussed in Refs. \cite{fisher,kat} and the associated phase
diagram is known exactly.  In the complex Boltzmann weight variable
$z=e^{-2K}$ (see below for notation) the phase boundaries consist of two
intersecting circles with a $z \to -z$ symmetry; more compactly, in the
variable $u=z^2$ which incorporates this symmetry, they are given by a
lima\c{c}on of Pascal \cite{chisq}.  Just as this generalization to
complex-temperature has yielded a deeper insight into the zero-field model, so
also it can shed light on the behavior of the model for nonzero field.
Accordingly, in this paper, we shall investigate the complex-temperature
properties of the 2D Ising model for nonzero external field.
Our methods include calculations of complex-temperature zeros of the
partition function and analyses of low-temperature series expansions.

   Although no one has solved the 2D Ising model in an arbitrary field,
Lee and Yang did succeed in solving exactly for the free energy and
magnetization for a particular manifold of values of $H$ depending on
the temperature $T$, given by $H = i (\pi/2) k_B T$ \cite{ly} (see also
\cite{yl}).
Although this is not a physical set of values, owing to the imaginary
value of $H$ and the resultant non-hermiticity of the Hamiltonian, this
model is nevertheless of considerable interest for the insight which it yields
into the properties of the Ising model in the
presence of a symmetry-breaking field.  By an extension of our analysis to
complex values of $H$, we are able to continuously connect the two known exact
solutions of the 2D Ising model, at $H=0$ \cite{ons, yang} and at
$H=i(\pi/2)k_BT$ \cite{ly}.

    Before proceeding, we mention some related work.  Rigorous results
on the behavior of the model for nonzero external field $H$
include the theorem that, for ferromagnetic (FM) spin-spin coupling, $J>0$,
the free energy $F(T,H)$ is an analytic function of temperature
\cite{yl}-\cite{lp}.  In the case of antiferromagnetic (AFM) coupling,
$J < 0$, this is not the case; there is a temperature $T_b(H)$ such that as $T$
decreases through $T_b(H)$, the system undergoes a (first-order) transition
 from FM to AFM long-range order.
Although there is no known exact expression for $T_b(H)$,
accurate numerical values are known (e.g., \cite{bdn} and references
therein). In addition to the works noted above \cite{fisher,kat}, papers
on complex-temperature singularities in the 2D square-lattice Ising model for
$H=0$ include Refs. \cite{g69}-\cite{egj}.  A useful connection between the
Lee-Yang solution for $\beta H=i\pi/2$ and a modified zero-field Ising model
with certain coupling(s)
shifted by $i\pi/2$ was discussed in Refs. \cite{merlini,lw}.  The
complex-temperature phase diagram for the Lee-Yang case
$\beta H = i\pi/2$ was
worked out in Ref. \cite{ih}, inlcuding exact results for the specific heat and
magnetization critical exponents and series analyses to determine the
susceptibility critical exponents at certain complex-temperature
singularities.  Complex-temperature zeros of the partition were studied for the
3D Ising model in a nonzero magnetic field in Ref. \cite{ipz}.
In contrast to the Ising model, a certain
superexchange model can be solved exactly in a field \cite{fsm}.  We have
studied the complex-temperature properties of this model for nonzero field and
will report the results elsewhere.  Whereas here we study complex-temperature
zeros of the partition function for real and certain complex values of
magnetic field, a different complexification is to study the zeros of $Z$ in
the complex magnetic field for physical temperature \cite{yl,ly}.  This led
the famous theorem \cite{yl,ly} that for ferromagnetic couplings the zeros
in the complex $e^{-2\beta H}$ plane lie on a circle and pinch the positive
real axis as the temperature decreases through its critical value; this work
also derived integral relations expressing thermodynamic quantities in
terms of integrals of densities of complex--field zeros.  Similar relations for
complex--temperature zeros were later discussed in Refs. \cite{fisher,abe}.

\section{General Properties}

\subsection{Model}

   Our notation is standard and follows that in our earlier works (e.g.,
\cite{chisq,ih}), so we review it only briefly. The (spin $1/2$,
isotropic, nearest-neighbor) Ising model on the square lattice
is defined by the partition function
\beq
Z = \sum_{\{\sigma_n\}} e^{-\beta {\cal H}}
\label{z}
\eeq
with the Hamiltonian
\beq
{\cal H} = -J \sum_{<nn'>} \sigma_n \sigma_{n'} - H \sum_n \sigma_n
\label{ham}
\eeq
where $\sigma_n = \pm 1$ are the $Z_2$ spin variables on each site $n$ of the
lattice, $\beta = (k_BT)^{-1}$, $J$ is the exchange constant,
$<n n'>$ denote nearest-neighbour sites,
and the units are defined such that the
magnetic moment which would multiply the $H\sum_n \sigma_n$ is unity.
(Hereafter, we shall use the term ``Ising model'' to denote this model
unless otherwise indicated.) For $H=0$, the symmetry group of the theory is
Z$_2$. We use the notation
\beq
K = \beta J \ , \quad h = \beta H
\label{kh}
\eeq
\beq
z=e^{-2K} \ , \quad u = z^2 = e^{-4K} \ , \quad v = \tanh K
\label{zuv}
\eeq
and
\beq
\mu=e^{-2h}
\label{mu}
\eeq
The reduced free energy per site is
$f = -\beta F = \lim_{N_s \to \infty} N_s^{-1} \ln Z$ in the thermodynamic
limit, where $N_s$ is the number of sites on the lattice. The
zero-field susceptibility is
$\chi =\frac{\partial M(H)}{\partial H}|_{H=0}$, where $M(H)$ denotes the
magnetization.  It is convenient to deal with the reduced quantity $\bar\chi =
\beta^{-1}\chi$.

A useful property is that the partition function $Z$ is a generalized
polynomial (with both negative and positive integral powers) in $u$ and
$\mu$.  (For a lattice with odd coordination number, $Z$ would be a generalized
polynomial in $z$ and $\mu$.)  On a finite lattice, for fixed $\mu$, $Z$ thus
has a certain set of zeros in the $u$ plane.
Experience with the zero-field model shows
that in the thermodynamic limit, these merge together to form curves (including
possible line segments) across which the free energy is non-analytic.  These
are the only singularities of $f$, except for the trivial singularities when
$|K| = \infty$ (which are isolated singularities and thus are not
important for the discussion of phase boundaries.)
Hence, the calculation of zeros of the partition function on
finite lattices serves as a valuable means by which to gain information about
the above continuous locus of points where the free energy is non-analytic in
the thermodynamic limit.\footnote{Parenthetically, we note that for anisotropic
spin-spin couplings, the complex-temperature zeros would merge to form areas
instead of curves in the thermodynamic limit \cite{aniso}.  Indeed, for the
(heteropolygonal) $4 \cdot 8^2$ lattice, even if the couplings are
isotropic, the zeros still form areas in this limit \cite{cmo}.}
This continuous locus of points includes the phase boundaries of the
complex-temperature phase diagram.  It may also include certain arcs or line
segments which protrude into and terminate in the interior of some phases and
hence do not separate any phases.

\subsection{Symmetries}
\label{sym}

   We record here some basic symmetries which will be used in our work.
First, because $Z$ is a generalized polynomial in $\mu$, in the analysis of
the phase diagram, it suffices, with no loss of
generality, to consider only the range
\beq
 -\frac{i\pi}{2} < Im(h) \le \frac{i\pi}{2}
\label{imhrange}
\eeq
 Second,
the summand of the partition function is invariant under the transformation
$h \to -h$, $\sigma_n \to -\sigma_n$.  The sign flip $h \to -h$ is equivalent
to the inversion map
\beq
\mu \to \frac{1}{\mu}
\label{muinversion}
\eeq
Hence, in considering nonzero real $h$, one may, with no loss of
generality, restrict to $h \ge 0$.  More generally, in considering complex $h$,
one may, with no loss of generality, restrict to the unit disk
\beq
   |\mu| \le 1
\label{mudisk}
\eeq
in
the $\mu$ plane.  We shall concentrate here on the range of $\mu$
values which connect the Onsager and Lee-Yang solutions of the model,
viz.,
\beq
-1 \le \mu \le 1
\label{muinterval}
\eeq
By the above symmetry, these values suffice to
describe the entire real line
\beq
 -\infty \le \mu \le \infty
\label{muline}
\eeq

   Third, concerning the symmetries of the locus of points across which the
free energy is non-analytic
and the associated complex-temperature phase diagram
(and, for finite lattices, the set of zeros of $Z$),
these are invariant under  $u \to u^*$, i.e. under reflection about the
horizontal, $Re(u)$ axis.  Fourth, if and
only if $\mu=\pm 1$, this locus of points is invariant under the inversion map
\beq
u \to \frac{1}{u}
\label{uinversion}
\eeq
This symmetry holds because of the bipartite nature of the square
lattice, and this mapping interchanges the uniform and staggered
magnetizations.  Thus, if one starts at a point in the FM phase, the mapping
takes one to a corresponding point in the AFM phase.  Since, in general, the
FM phase occupies a region in the neighborhood of the origin in the $u$ (or
$z$) plane, this shows that the AFM phase will be the outermost phase,
extending to complex infinity in these two respective planes.
This result is obvious for $h=0$ and can be seen easily for
the case $\mu=-1$ from the relation \cite{merlini,lw}
connecting this theory to a zero-field Ising model with certain couplings
shifted by $i\pi/2$, as discussed in \cite{ih}.  Although this inversion
symmetry does not hold for $\mu \ne \pm 1$, one can still make the following
statement.  Let us write $u$ in polar form, $u = \rho_u e^{i\theta_u}$.  Then
\beq
K = -\frac{1}{4}\ln u = -\frac{1}{4}\Bigl ( \ln \rho_u + i\theta_u
+ 2\pi i n \Bigr )
\label{keq}
\eeq
where $n$ indexes the Riemann sheet of the logarithm and will not be important
here (we take it to be $n=0$).  The usual limit of infinitely strong
antiferromagnetic spin-spin exchange coupling $K= \beta J \to -\infty$
corresponds to $u$ going to infinity along the positive real axis.  However,
eq. (\ref{keq}) shows that as $|u| \to \infty$ along any direction, not
necessarily $\theta_u=0$, $K$ corresponds to an infinitely strong AFM coupling,
since the term $-(1/4)i\theta_u$ is finite and becomes negligible in this
limit.  It follows that for arbitrary fixed $\mu$, if an AFM phase exists at
all (it is absent, e.g., for the isotropic Ising model on the triangular
and Kagom\'e lattices),
then it extends outward in all directions to complex infinity in the
$u$ or $z$ plane.  This general property was evident in our earlier studies of
the complex-temperature phase diagrams for the zero-field Ising model on the
square, triangular, and honeycomb lattices \cite{chisq,chitri}, as well as
the heteropolygonal $3 \cdot 12^2$ and $4 \cdot 8^2$ lattices \cite{cmo}.

    Finally, we can show a close connection between the properties of the
theory in the vicinity of the origin in the $\mu$ plane by a similar argument.
Small negative $\mu$, i.e. $\mu \to 0^-$, is
equivalent to small positive $\mu$, $\mu \to 0^+$, which, in turn, corresponds
to infinitely strong external field, $h \to \infty$.  These results are easily
seen as follows. Let us write $\mu = \rho_\mu e^{i\theta_\mu}$.  Then from the
definition $\mu=e^{-2h}$,
\beq
h = -\frac{1}{2}\ln \mu = -\frac{1}{2}\Bigl ( \ln \rho_\mu + i\theta_\mu
+ 2\pi i n \Bigr )
\label{heq}
\eeq
  As $\rho_\mu \to 0$, $h \to \infty$.  In this
limit, the term involving the argument $\theta_\mu$ has a negligible effect.
(This actually yields a
stronger result than we need here, viz., that as $\rho_\mu \to 0$ along
any direction, not just $\theta_\mu=0$ or $\pi$, this direction becomes
asymptotically unimportant,
and the effect is like a physical uniform field.) This implies that in the
limit as $\mu \to 0$ (along any direction), the
complex-temperature phase diagram in the $u$ plane consists only of the FM
phase, with the AFM phase frozen out.  As a consequence of the $h \to -h$, $\mu
\to 1/\mu$ symmetry mentioned above, this also applies (with $M \to -M$) to the
limit $|\mu| \to \infty$ along any direction.

\section{Complex-Temperature Properties for Solved Cases}

\subsection{$h=0$ ($\mu=1$)}

    In order to achieve a better understanding of the results that we shall
obtain concerning the complex-temperature phase diagram
for the interval $-1 < \mu < 1$ (or equivalently, the union
$\{ -\infty \le \mu < -1 \} \quad \bigcup \quad
\{ 1 < \mu \le \infty \}$) where the model
has not been solved exactly, it is useful to discuss this phase diagram for
the two cases where it is known precisely, viz., $\mu=\pm 1$.  As part of
Fig. 1, we show the complex-temperature phase diagram for $H=0$  ($\mu=1$).
The phase boundaries in the $u$ plane form a lima\c{c}on \cite{chisq},
given by
\beq
Re(u) = 1 + 2^{3/2}\cos \omega + 2 \cos 2\omega
\label{reu}
\eeq
\beq
Im(u) = 2^{3/2}\sin \omega + 2 \sin 2\omega
\label{limacon}
\eeq
traced out completely for $0 \le \omega < 2\pi$ (see Fig. 1).
The complex-temperature extensions of the physical
phases with spontaneous Z$_2$ symmetry breaking
via ferromagnetic and antiferromagnetic long range order lie,
respectively, within the inner branch of the lima\c{c}on and outside the outer
branch of this lima\c{c}on.  The complex-temperature extension of the
Z$_2$--symmetric, paramagnetic (PM) phase lies between the inner and outer
branches of the lima\c{c}on.  Note that the FM, AFM and (two wedges of the) PM
phase are all contiguous at the point $u=-1 \equiv u_s$.  Recall that the
physical critical point separating the PM and FM phases is $u_c=3-2\sqrt{2}
=0.171573...$,
and, in accordance with the $u \to 1/u$ symmetry noted above, the corresponding
critical point separating the PM and AFM phase is
$u=1/u_c=3+2\sqrt{2}=5.82843...$.  Henceforth, we shall generally refer to
these phases simply as FM, AFM, and PM, with the qualifier
``complex-temperature extension'' being understood.  The corresponding phase
boundaries in the $z$ plane consist of the intersecting circles
$z=\pm 1 + 2^{1/2}e^{i\theta}$, $0 \le \theta < 2\pi$ \cite{fisher,kat}.  (In
this variable, there are actually four phases: FM, AFM, PM, and O, where
the O phase has no overlap with any physical phase \cite{chisq}.)

   The spontaneous magnetization is
\beq
M(u,h=0)=\frac{(1+u)^{1/4}[(1-u/u_c)(1-u_cu)]^{1/8}}{(1-u)^{1/2}}
\label{m}
\eeq
in the physical FM phase \cite{yang},
and this expression holds, by analytic continuation, throughout the
complex-temperature extension of the FM phase ($M=0$ elsewhere).
Thus, $M$ vanishes continuously
both at the physical critical point $u=u_c$ and at the point $u=-1\equiv
u_s$.  $M$ vanishes discontinuously elsewhere along the boundary of the
(complex-temperature extension of the) FM phase.
As discussed in Ref. \cite{chisq}, the apparent singularities at
$u=1/u_c$ and $u=1$ do not actually occur since these points lie outside the
FM phase where the above expression
applies; indeed, $M$ vanishes identically in the vicinity of $u=1/u_c$ and
$u=1$.  By a well-known symmetry, the staggered magnetization is given by
$M_{st}(y) = M(u \to y)$, where $y=1/u$, in the AFM phase and is zero
elsewhere. In addition to its
well-known divergence at the physical critical point, the susceptibility
$\bar\chi$ diverges at $u=-1$ with exponent inferred from series analysis to be
$\gamma_s'=3/2$ \cite{egj,chisq}.  Ref. \cite{chisq} obtained the relation
\beq
\gamma_s = 2(\gamma-1)
\label{gammasrel}
\eeq
which explains
the value of $\gamma_s'$ in terms of that of the usual exponent $\gamma=7/4$.

\subsection{$h=i\pi/2$ \quad ($\mu=-1$)}

   In Ref. \cite{ih} we determined the locus of points across which the free
energy is non-analytic, and the corresponding complex-temperature phase
diagram, for the Lee-Yang solution at $\mu=-1$.  These consist of the union of
the unit circle and a certain line segment on the negative real $u$ axis:
\beq
\{ \ u = e^{i\theta} \ , \ \theta \in [0,2\pi) \ \} \qquad \bigcup \qquad
\{ \ 1/u_e \le u \le u_e \ \}
\label{usingly}
\eeq
where
\beq
u_e=-(3-2\sqrt{2})
\label{ue}
\eeq
(which is also equal to minus the value of the critical point $u_c$ in the
$h=0$ model.)  Since the nonzero value of $H$ explicitly breaks the Z$_2$
symmetry, there is no PM phase.  The FM and AFM phases occupy the interior
and exterior of the unit circle, respectively.
In contrast to the $h=0$ case, here a subset of the continuous locus of points
across which $f$ is non-analytic, viz., the line segment in (\ref{usingly}),
does not completely separate any phases, but instead protrudes into the FM
phase and AFM phases, terminating in the respective endpoints $u=u_e$ and
$u=1/u_e$. The magnetization is \cite{ly}
\beq
M(u,h=i\pi/2) = \frac{(1+u)^{1/2}}{(1-u)^{1/4}[(1-u/u_e)(1-u_eu)]^{1/8}}
\label{mh}
\eeq
within the FM phase (and zero elsewhere).  As is clear from (\ref{mh}), this
vanishes continuously at
$u=-1$ with exponent $\beta_s = 1/2$, and diverges at $u=u_e$ with exponent
$\beta_e = -1/8$ and at $u=1$ with exponent $\beta_1 = -1/4$.
Elsewhere on the boundary of the FM phase,
i.e., the unit circle in the $u$ plane, $M$ vanishes discontinously. Note that
the apparent divergence at the point $u=1/u_e$ does not actually
occur, since this is outside of the complex-temperature FM phase, where the
above analytic continuation is valid.  From analyses of low-temperature
series, we concluded that $\bar\chi$ has divergent singularities (i)
at $u=u_e$ with exponent $\gamma_e'=5/4$, (ii) at $u=1$, with
exponent $\gamma_1'=5/2$, and (iii) at $u=u_s=-1$, with exponent
$\gamma_s'=1$ \cite{ih}. (The actual values obtained from series analysis were
$\gamma_e'=1.25 \pm 0.01$, $\gamma_1'=2.50 \pm 0.01$, and $\gamma_s'=1.00 \pm
0.08$.)

\subsection{Partition Function Zeros for Solved Cases}

\subsubsection{$\mu=1$}

    Since one of our two main methods of gaining information about the
complex-temperature phase diagram of the model for $\mu \ne \pm 1$
is the use of complex-temperature zeros of $Z$, it is important to see
how accurate this method is for the two cases where one knows the
phase diagram exactly, viz., $\mu=\pm 1$.  That is, we wish to ascertain
how well, for finite lattices with various boundary
conditions, the pattern of zeros resembles the locus of points across which
$f$ is non-analytic in the thermodynamic limit.

   In order to compute the zeros, we calculate $Z$ for finite lattices with
specified boundary conditions (BC's).  We have done this by means of a
transfer matrix method \cite{tm}, and have used both periodic and helical
boundary conditions (respectively PBC, HBC). We recall that in order to avoid
frustration of (short- or long-range) AFM ordering, it is necessary and
sufficient that the lengths $L_1$ and $L_2$ of the lattice be (i) both even for
PBC and (i) one even, the other odd, for HBC.  We have incorporated this
restriction in our work.

     In Fig. 1 we show the comparison for the $h=0$ case.  We have found that
for similar-size lattices, the calculation with helical boundary conditions
yields zeros which lie generally slightly closer to the lima\c{c}on than the
calculation with periodic boundary conditions.  Accordingly, we show in Fig. 1
the zeros for a $7 \times 8$ lattice with helical boundary conditions.
In the vicinity of the FM-PM and PM-AFM critical points the
density of zeros calculated on the finite lattice decreases as one approaches
the real axis, in accord with the exact result \cite{fisher,abe} that
in the thermodynamic limit this density $g$ vanishes, in our notation, like
$g(u,\mu=1) \sim |(1-u/u_c)(1-uu_c)|^{1-\alpha}$ as $u \to u_c$ or $u \to
1/u_c$ along the curves of zeros, where $\alpha=\alpha'$ is the specific heat
exponent, for both of these critical points.  As one approaches the singular
point $u=-1$ from within the FM, AFM, or PM phases, the specific heat $C$
has logarithmic divergences \cite{chisq}, i.e., $ \alpha_s=\alpha_s'=0$
Together with a generalization of the above relation for $g$, viz.,
\beq
g \sim |1-u/u_{sing.}|^{1-\alpha_{sing.}'} \quad as \ \ u \to u_{sing.}
\label{gsing}
\eeq
along the curve(s) where $g$ has nonzero support,
this implies that $g(u,\mu=1) \sim |1+u| \to 0$ as $u \to -1$
along the curves of zeros comprising the complex-temperature phase boundaries.
The behavior of the zeros in Fig. 1 is evidently consistent with this exact
result.  Indeed, one observes that the zeros avoid the region $u=-1$.  A
similar tendency of the zeros not only to become less dense but also to
deviate from the lima\c{c}on can be seen in the vicinity of the PM-AFM critical
point, where the last complex-conjugate pair of zeros occur significantly
within the curve.  (By the $u \to 1/u$ symmetry, this is equivalent to the
inverse deviation of the last complex-conjugate pair of zeros near the
FM-PM critical point.)  The fact that there are no
zeros on the positive real $u$ axis is, of course, also implied by the property
that for $H=0$, the partition function for a finite lattice is a generalized
polynomial in $u$ with positive coefficients.

\subsubsection{$\mu=-1$}

   We have performed a similar comparison for $\mu=-1$ and show a plot of
the zeros in Fig. 2, again for a $7 \times 8$ lattice with helical boundary
conditions.\footnote{To save space, we truncate the plot on the left at
$Re(u)=-2$; there are four zeros lying on the negative real axis, and a pair
lying near this axis, to the left of this point, which are not shown.  However,
the plot as shown contains all the information, since these six zeros are just
the inverses, by the $u \to 1/u$ symmetry, of the six which are shown, lying on
or near the negative real axis within the unit circle.}
One sees that all of the zeros
in the right-hand half plane, and several other zeros, lie exactly
(to within numerical accuracy) on the unit circle.  The zero at $u=1$ has
multiplicity $N_s/2$ ($=28$ here), consistent with the fact that in the
thermodynamic limit, the (reduced) free energy $f$ contains an additive term
$(1/2)\ln(1-u)$ (see Ref. \cite{ly} and eq. (2.29) of
Ref. \cite{ih})\footnote{As discussed in Ref. \cite{ih}, in addition to this
additive term there is also a subdominant contribution to the singularity in
$f$ at $u=1$ arising from the vanishing of the argument of the logarithm in
eq. (2.29) of that paper.} corresponding to a delta function in the density of
zeros, $g$, at $u=1$.  Because of
this delta function singularity in $g$, the relation
expressing the specific heat as an integral of $g$ over the region where it has
nonzero support receives an additive contribution.  In turn, this has the
effect of modifying the relation (\ref{gsing})
connecting $g$ with the critical
exponent in the specific heat at a singular point $u_{sing.}$,
Hence, the fact, shown in Ref. \cite{ih}, that the specific heat $C$
has a finite non-analyticity ($\alpha_1'=0$) at $u=1$ as this
point is approached from the FM or AFM phases does not conflict with the
non-vanishing of $g$ at $u=1$.  On the negative real axis, there are zeros
lying on (or near) the line segment in (\ref{usingly}).  As one
can see from Fig. 2, near the inner (and by the $u \to 1/u$ symmetry, also
the outer) endpoint of the line segment, the density of zeros
approaches a nonzero constant.  This finite-lattice feature is in good
agreement with the infinite-lattice relation (\ref{gsing}) between $g$ and
the specific heat singularity, together with the exact
exponent which we found earlier \cite{ih},
$\alpha_e'=1$, at $u=u_e$ and $u=1/u_e$.  Proceeding next to the intersection
point at $u=-1$, we observe that our exact result that the specific heat
has a finite non-analyticity ($\alpha_s'=0$) at $u=-1$ as this point is
approached from either the FM or AFM phases \cite{ih} implies that in the
thermodynamic limit, $g(u,\mu=-1) \sim |1+u| \to 0$ as
one approaches this point along the unit circle or the negative real axis.
This is consistent with the results in Fig. 2.  (It happens that for the $7
\times 8$ lattice with HBC there is a zero precisely at $u=-1$, but this does
not occur for other lattice sizes and is not a general feature.)  Finally, we
note a contrast with the $h=0$ case: there, $Z$ was a generalized
polynomial in $u$ with positive coefficients, whereas here the coefficients
are not all positive, so that zeros can, and indeed do, occur on the real
$u$ axis.   One quantitative measure of how well the zeros on the finite
lattices approach the locus of zeros on the infinite lattice is provided by how
close the innermost zero on the negative $Re(u)$
axis is to the infinite-lattice result, $u_e=-(3-2\sqrt{2})$.  We find that the
position of this zero is (i) $u=-0.1822$ on an $8 \times 8$ lattice with PBC,
and (ii) $u=-0.1831$ on a $7 \times 8$ lattice with HBC.  These points are
respectively 6.2 \% and 6.5 \% farther away from the origin than $u_e$.

\section{Complex-Temperature Properties $0 < h \le \infty$ \quad
($0 \le \mu < 1$)}

\subsection{Information from Zeros of $Z$}

  Having these comparisons with exactly solvable cases as background, we now
proceed to investigate the complex-temperature phase diagram and associated
singularities for the Ising model in a nonzero magnetic field, i.e., $0 < H \le
\infty$, or equivalently, $0 \le \mu < 1$.  Since any nonzero magnetic field
explicitly breaks the Z$_2$ symmetry of the Hamiltonian, the Z$_2$-symmetric,
PM phase immediately disappears.   We recall from section (\ref{sym})
that $H$ can be taken positive without loss of generality, and that the
results for the above interval in $\mu$ also describe the behavior for
 $1 < \mu \le \infty$.  Of the four symmetries in
section (\ref{sym}), the $u \to 1/u$ symmetry of the phase
diagram (with corresponding interchange of FM and AFM phases) ceases to hold if
$\mu \ne \pm 1$.  One may inquire how the complex-temperature phase boundary
changes as $H$ is increased from zero.  One knows rigorously that for $h
\ne 0$, it is possible to analytically continue from $u=0$ out along the
positive real $u$ axis through the point $u_c$ (which is no longer a critical
point), all the way to the vicinity of $u=1$ ($K=0$).  This means that the
region which was previously, for $H=0$, the PM phase and its
complex-temperature extension, immediately becomes part of the FM phase and its
complex-temperature extension, respectively.  This also means that
for any nonzero $h$, the inner branch of the lima\c{c}on which formed the
phase boundary between the FM and PM phases for $H=0$ breaks apart at $u=u_c$
into two separate arcs, leaving an opening through which the analytic
continuation mentioned above can be performed.

   One also knows that if $J < 0$,
then for a fixed value of $H$, at sufficiently low $T$, the system
will still exhibit long-range AFM ordering.  Equivalently, for fixed $h$ and
hence $\mu$, and sufficiently large $u > 1$ (i.e. sufficiently large negative
$K$), there will be AFM ordering.  Thus, for any finite $h$, there is still a
non-analytic boundary
separating the physical and CT-extension of the FM phase from the physical and
CT extension of the AFM phase.  Since the nonzero $H$ biases the system toward
FM ordering and hence against AFM ordering, it follows that the singular point
which separates the physical FM and AFM phases, $u_{_{FM,AFM}}$,
increases as $H$ increases (starting from the value
$\lim_{H \to 0} u_{_{FM,AFM}} = 1/u_c$).  As $u$ increases through the value
$u_{_{FM,AFM}}$, the system undergoes a first-order transition at which $M$
jumps discontinuously to zero.  One expects that generic points on the border
between the (complex-temperature) FM and AFM phases are also associated
with a first--order phase transition.
Given the biasing effect of nonzero $H$,
one expects the complex extension of the FM phase to increase as $H$
increases.  Because of the symmetry under $h \to -h$ noted above,
it follows that the rightward shift in $u_{_{FM,AFM}}$ as a function of $H$ is
independent of the sign of $H$.
An additional general statement is that for all $H$ (zero or not), the
(complex-temperature extensions of the) FM and AFM phases must be completely
separated by a phase boundary.

   In Fig. 3 we show a combined plot of the CT zeros of $Z$, calculated for the
same lattice ($7 \times 8$ with helical boundary conditions) as before,
for a sequence of increasing values of $h$, and in Fig. 4 we show plots of
these zeros for certain fixed values of $h$.  There are several
interesting features which we can discern from these calculations.
The results are in accord with the expectations based on rigorous
arguments noted above.  As $h$ increases above zero ($\mu$ decreases
below 1), the right hand end of the inner branch of the lima\c{c}on which
formerly crossed the real $u$ axis at $u_c$ immediately breaks at $u=u_c$,
rendering it possible to analytically continue from $u=0$ along the positive
real axis outward past $u=0$ all the way to the vicinity of $1/u_c$.  This
breaking yields two complex-conjugate arcs, which retract from the real $u$
axis as $h$ increases.  Scaling arguments imply that in the neighborhood of
$h=0$, the endpoint of the
arc in the upper half plane should move in the direction \cite{ipz}
\beq
\theta_{ae} = \frac{\pi}{2 \beta \delta}
\label{thetaae}
\eeq
For the 2D Ising model, inserting the exactly known critical exponents
$\beta=1/8$ and $\delta=15$, it follows that
\beq
\theta_{ae,2D}=\frac{4 \pi}{15} = 48^\circ
\label{thetaae2d}
\eeq
One can see roughly, and we have confirmed accurately that this is in agreement
with the motion of the zero nearest to $u_c$ for small $h$.  The direction of
motion away from the origin is different for the zeros on the inner
arcs.  Related to this, as $h$ increases, the upper arc becomes oriented in a
more northwest to southeast, rather than west to east, direction.

    A notable feature that one sees in each of the graphs in Fig. 4 is that
there is a sharp increase in the density of zeros as one approaches the
endpoints of these arcs.  This suggests the possibility that in the
thermodynamic limit the density $g$ might actually diverge as
one approaches these endpoints along the arcs.  Below, we
shall find support for this conclusion from our analysis of
low-temperature series, which yields a value for the specific heat exponent
$\alpha_{ae}' > 1$ at these endpoints; together with the relation
(\ref{gsing}), such a value of $\alpha_{ae}'$ implies that $g$ diverges at the
arc endpoints.  Note also, of course, that $\alpha_{ae}' > 1$ also means
that the internal energy diverges at these endpoints.  In this context, we
recall that in our
study of the complex-temperature singularities of the Ising model on the
honeycomb lattice \cite{chitri}, we found the exact result that at
the point $z=-1 \equiv z_\ell$, as approached from either the FM or AFM
phases, the specific heat diverges with an exponent $\alpha_\ell'=2$.

   In accord with the argument given above, as $h$ increases,
$u_{_{FM,AFM}}$ increases, and, more generally, one sees
that in the right-hand half plane, the zeros on what was
the outer branch of the lima\c{c}on for $h=0$ move outward.
The directions of motion of these
zeros are approximately radially outward and do not change significantly as $h$
is increased.  Of course, the fact that the zeros on what were, for $h=0$, the
inner and outer branches of the lima\c{c}on both move outward is a striking
manifestation of the loss of the $u \to 1/u$ symmetry which held for $h=0$,
since the latter symmetry implied that these zeros were inverses of each other,
(so that if the inner branch moved out, the outer branch would have had to
move in).  From the comparison of the zeros calculated on finite lattices with
the exact phase boundary for $H=0$, we saw that the pair of zeros
farthest to the right lies within the outer branch of the lima\c{c}on. One
thus anticipates that this may also be true for nonzero $H$.  Numerical
estimates of $K_b=J/(k_BT_b(H))$, expressed in terms of the $u$ variable, are
$u_{_{FM,AFM}}=5.90909(9)$, 6.34036(9), and 7.99333(10) for
$h=0.2$, 0.5, and 1.0, respectively, where the number in brackets is the
uncertainty in the last digit \cite{bdn}.  If we draw
a curve through the zeros on the right-hand side of the phase diagram,
and then compare the point at which they cross the real axis with these values,
we see that, indeed, the finite lattice calculations yield a slight
underestimate of this point for nonzero $H$, as they did for $H=0$.

 The motion of the zeros in the left-hand half-plane is more complicated.  One
sees that, while the zeros do move generally outward from the origin as $H$
increases, the directions of motion of a number of zeros undergo significant
(and non-monotonic) changes.
We note that, owing to the loss of the inversion symmetry for $\mu \ne \pm 1$,
the self-inverse point $u=-1$ loses the special significance which it had for
these two values, for which it was an intersection point on the respective
phase boundaries.

   From these calculations, we may draw some further inferences about the phase
diagram in the thermodynamic limit.  As $h$ increases, the boundary between the
FM and AFM phases on the right-hand side of the phase diagram
moves outward from the outer branch of the lima\c{c}on.  On the left-hand side,
the intersection point at $u=-1$ is replaced by an outer boundary to the left
of this point, which shows some concavity but moves farther to the left as $h$
increases. Recall that for $h=0$, the inner and outer branches of the
lima\c{c}on cross in a perpendicular manner at this point \cite{chisq} (see
Fig. 1).  Our calculations suggest that it is likely that as $h$ increases
from 0, the curves which, for $h=0$, were the inner and outer branches of the
lima\c{c}on, no longer intersect at $u=-1$; instead, the upper and lower parts
of the former outer branch join smoothly to each other on the left, and cross
the real $u$ axis in a vertical manner rather than as perpendicular NW--SE and
SW--NE curves.  Our results also suggest that the two complex-conjugate
arcs which, for $h=0$, comprised the connected inner branch of the
lima\c{c}on, no longer cross the negative real $u$ axis at $u=-1$ but instead
retract from this axis.  It is possible that they join the outer phase
boundary (at complex-conjugate points).   As $h$ increases more, our calculated
zeros indicate
that these complex-conjugate arcs move progressively farther from
the real axis.  As $h \to \infty$, the
boundary separarating the FM and AFM phases moves outward to
complex infinity in the $u$ plane, so that in this limit, the entire plane is
occupied by the FM phase.  We shall next obtain further
information about the endpoints of these arcs from a series analysis.

\subsection{Results from Series Analysis}

       In order to investigate the complex-temperature singularities for $-1
< \mu < 1$, we shall make use of the low-temperature, high-field series
expansion for the free energy of the Ising model on the
square lattice \cite{tlow1}--\cite{be}.
The partition function $Z$ can be written
as $Z = e^{N_s(2K+h)}Z_r$, and equivalently, the reduced free energy can be
written as $f=2K+h+f_r$, where $f_r=\lim_{N_s \to \infty}N_{s}^{-1}\ln Z_r$.
In Ref. \cite{be}, Baxter and Enting calculated the low-temperature expansion
of the quantity $\kappa = e^{f_r} = \lim_{N_s \to \infty} Z_r^{1/N_s}$
to $O(u^{23})$ (with the coefficients of the powers of $u$ being exact
polynomials in $\mu$).  This expansion has the form
\beq
\kappa = 1 + \sum_{n=2}^{\infty}\sum_{m} a_{n,m}u^n \mu^m
\label{kappa}
\eeq
where $j \le m \le j^2$ for $n=2j$ and $j \le m \le j(j-1)$ for $n=2j-1$.  From
this series, we calculate, for each value of $\mu$, resultant series for the
specific heat $C$, magnetization $M$, and susceptibility $\bar\chi$. We then
analyze these using dlog Pad\'e and differential approximants (abbreviated PA
and DA; for a recent review, see \cite{tonyg}).
Our notation for differential approximants follows that in Ref. \cite{tonyg}
and our implementation is the same as that in Ref. \cite{ih}; in particular,
we use unbiased, first-order differential approximants.

    We find from both the PA and DA analyis that for $0 < \mu < 1$, the
series indicate a complex-conjugate pair of singularities at positions which
are in very good agreement with
the locations of the innermost zeros on the inner
arcs.  We denote these arc endpoints, $u_{ae}$ and $u_{ae}^*$.  Table 1 shows
the comparison of the values of the arc endpoint $u_{ae}$ as obtained from the
complex-temperature zeros on finite lattices and from the analysis of the
low-temperature series.  The fractional differences in the
positions are about 2 \%.

\begin{table}
\begin{center}
\begin{tabular}{|c|c|c|c|c|c|c|} \hline \hline & & & & & & \\
$h$ & $\mu$ & $u_{ae,Z}$ & $u_{ae,ser.}$ & $\alpha_{ae}'$ &
$\beta_{ae}$ & $\gamma_{ae}'$ \\
& & & & & & \\
\hline \hline
0.8  & 0.202 &  $0.313+0.663i$ & $0.3203(6)+0.6503(6) i$ & 1.175(7)
& $-0.18(5)$ & 1.195(10) \\ \hline
1.0  & 0.135 &  $0.320+0.857i$ & $0.3265(5)+0.8426(6) i$ & 1.177(8)
& $-0.18(5)$ & 1.20(2) \\ \hline
1.2  & 0.0907 & $0.324+1.086i$ & $0.3306(5)+1.0685(6) i$ & 1.18(2)
& $-0.18(5)$ & 1.19(2)  \\ \hline
\hline
\end{tabular}
\end{center}
\caption{Values of the arc endpoint, $u_{ae}$, for several values of $h$,
 from (i) calculation of zeros of $Z$ on finite lattices, yielding $u_{ae,Z}$,
and (ii) analyses of series for $C$, $M$ and $\bar\chi$, yielding
$u_{ae,ser.}$.  Table also gives values of the exponents $\alpha_{ae}'$,
$\beta_{ae}$, and $\gamma_{ae}'$ at $u_{ae}$ as determined from analysis of
low-temperature series using dlog Pad\'e and differential approximants.
The values of $u_{ae,ser.}$ are quoted to an accuracy in accord with the
agreement between the different series and methods of analysis.  The values for
$u_{ae,Z}$ are quoted to an accuracy reflecting the differences in values
with different boundary conditions. The numbers in brackets
denote the uncertainties in the last digit of a given entry.}
\label{table1}
\end{table}

    We have made a preliminary study of the singularities at these arc
endpoints.  Since the arc endpoints arise abruptly with the breaking of the
inner branch of the lima\c{c}on for any $h$, regardless of how small, one
anticipates a similarly abrupt change in exponents.  Indeed, in our previous
studies of complex-temperature singularities (e.g. Refs.
\cite{chitri,cmo}), using exact results, we found that $M$ always diverges
at endpoints of arcs protruding into the FM phase.  Hence, we expect (and will
verify) that $M$ diverges at the arc endpoints in the present case, so that
$\beta$ jumps discontinuously from the value $\beta=1/8$ for $h=0$
to negative value(s) for nonzero $h$.  We assume the usual leading
singular forms
\beq
C \sim A_C(1-u/u_{ae})^{-\alpha_{ae}'}
\label{csingc}
\eeq
\beq
M \sim A_M(1-u/u_{ae})^{\beta_{ae}}
\label{csingm}
\eeq
\beq
\bar\chi \sim A_\chi(1-u/u_{ae})^{-\gamma_{ae}'}
\label{csingchi}
\eeq
(We do not consider confluent singularities here; at least for $H=0$, they
are unimportant for the 2D Ising model, a fact used also in our previous study
of complex-temperature singularities in this case \cite{chisq}.).
We analyze the singularities for three values of $h$ which are chosen to be
sufficiently large that the arc endpoints are reasonably well separated from
each other, viz., $h=0.8$, 1.0, and 1.2.  We obtain the results shown in Table
1.  In the case of $\alpha_{ae}'$ and $\gamma_{ae}'$, the dlog Pad\'e and
differential approximants yielded values in very good agreement with each
other.  In the case of $\beta_{ae}$, the PA's gave somewhat larger values than
the DA's; we list values which are more heavily weighted by the (presumably
more accurate) DA's, with estimated errors which are increased to reflect the
PA results.  As is evident from Table 1, for the values of $h$ which we have
considered, these exponents are consistent with being independent of the
value of $h$ in this range.  The values of the exponents are also consistent
with the equality $\alpha_{ae}'+2\beta_{ae}+\gamma_{ae}'=2$.  The fact that
$\beta_{ae}$ is negative means that $M$ diverges at the arc endpoints; this is
the same behavior that we found earlier in our studies of complex-temperature
singularities on various lattices, using the respective exact expressions for
the magnetization (e.g. see the summary in Table 2 of Ref. \cite{cmo}).  Our
results also indicate that $\alpha_{ae}' > 1$.  As noted above, in conjunction
with the relation (\ref{gsing}), this implies that in the thermodynamic limit
the density $g$ of zeros of $Z$ diverges as one moves along the arcs and
approaches their endpoints.  The behavior of the zeros calculated on finite
lattices already suggests this infinite-lattice result.

\section{Phase Diagram for $-1 \le \mu \le 0$}

\subsection{Zeros of $Z$}

   This range of $\mu$ is not physical, in contrast to the range $0 \le \mu \le
1$ discussed above.   Nevertheless, it is of interest because by determining
how the phase diagram changes as $\mu$ decreases through this range, we can
complete the connection between the exact solutions at $\mu=1$ and $-1$.

   For the discussion of the complex-temperature zeros of $Z$, just as in the
previous section, we
started with the exactly known case $\mu=1$ and moved away from it by
decreasing $\mu$,  so also it is convenient here to start with the exactly
known phase diagram for $\mu=-1$ and move away from it by decreasing
the magnitude of $\mu$ toward 0.  In Fig. 5 we show a plot of the paths of
individual zeros for a sequence of values of $h$, and in Fig. 6 we show plots
of zeros for fixed values of $h$.  These results have a number of salient
features.  First, as $\mu$ moves from $-1$ toward $0^-$, the zeros which were
on the unit circle move approximately radially outward to larger
values of $|u|$. They maintain a roughly circular form.  The zeros which had
been on or close to the negative real axis and had corresponded to what, in the
thermodynamic limit, was the line segment in (\ref{usingly})
extending between $1/u_e$ and $u_e$
for $\mu=-1$, move gradually outward, away from the origin.
Meanwhile, for any value of $\mu$ greater than $-1$,
there abruptly appears a new set of zeros on the positive real axis.
Our calculations of zeros on
finite lattices suggest the inference that in the thermodynamic limit, these
form a finite line segment of singular points starting initially at
$u=1$ for $\mu = -1 + \epsilon$, $\epsilon \to 0$ (and extending rightward).
As $\mu$ increases from $-1$,
the right-hand phase boundary between the FM and AFM phases moves outward
sufficiently quickly that, as one can see in Fig. 6, it lies to the right of
the new line segment of zeros on the positive real $u$ axis.
We shall denote the respective endpoints, in the thermodynamic limit, of the
left-hand and right-hand line segments of zeros as $u_{lhe}$ and $u_{rhe}$.
A striking
feature of the plots is that the density of zeros increases sharply as one
moves rightward toward the inner end of the line segment of
zeros on the negative real $u$ axis, and also as one moves leftward toward
the inner end of the line segment of zeros on the positive real axis.  This is
quite similar to what we observed for the density of zeros near the arc
endpoints in the FM phase for $0 \le \mu < 1$.  Again, it suggests the
inference that in the infinite-lattice limit, the density of zeros $g$
diverges at these endpoints $u_{lhe}$ and $u_{rhe}$.  Our series analysis
below support this inference.
The results in Fig. 6 also show that as $\mu$ moves toward 0 through
the interval $-1 < \mu < 0$, the endpoints of the line segments, $u_{lhe}$ and
$u_{rhe}$ move slowly away from the origin.  As $\mu \to 0^-$, the boundary
separating
the FM and AFM phases moves outward to complex infinity, so that in this limit,
only the FM phase remains.  This can be understood on
general grounds, as we have discussed above.

\subsection{Series Analysis}

    From our analysis of the low-temperature series for $C$, $M$, and
$\bar\chi$, we have obtained very good agreement with the positions of the
inner endpoints of the left and right line segments, $u_{_{lhe}}$ and
$u_{_{rhe}}$.
Table 2 displays this comparison.  Of course, one cannot use the series to
locate the outer endpoints of the left and right line segments since these lie
outside the FM phase where these series are applicable.

\begin{table}
\begin{center}
\begin{tabular}{|c|c|c|c|c|c|} \hline \hline & & & & & \\
$h_r$ & $\mu$ & $u_{_{lhe,Z}}$ & $u_{_{lhe,ser.}}$ & $u_{_{rhe,Z}}$
& $u_{_{rhe,ser.}}$ \\
& & & & & \\
\hline \hline
0.347 & $-0.5$ &  $-0.2976$ & $-0.2870(5)$ & 1.047 & 1.053(15) \\ \hline
0.805 & $-0.2$ &  $-0.5467$ & $-0.5314(8)$ & 1.246 & 1.245(10) \\ \hline
1.151 & $-0.1$ &  $-0.8471$ & $-0.8267(8)$ & 1.529 & 1.522(10)  \\ \hline
\hline
\end{tabular}
\end{center}
\caption{Values of the inner endpoints of the left and right line segments
for negative
$\mu$, with $h=-(1/2)\ln \mu = h_r + i\pi/2$.  Entries in the columns denoted
$u_{lhe,Z}$ and $u_{_{rhe,Z}}$ are the innermost zeros on the left- and
right-hand
line segments, calculated on a $7 \times 8$ lattice with helical boundary
conditions.  Entries in the columns denoted $u_{lhe,ser.}$ and $u_{rhe,ser.}$
are the positions of the inner endpoints of the left and right line segments as
obtained from the analyses of low-temperature series for $C$, $M$, and
$\bar\chi$.  Accuracies listed are chosen as in Table 1.}
\label{table2}
\end{table}

Since
\beq
\lim_{\mu \to -1} u_{lhe} = u_e
\label{uelim}
\eeq
and since the exponents are known for $\mu=-1$ as \cite{ih}
$\alpha_e'=1$, $\beta_e=-1/8$ (exact), and $\gamma_e'=5/4$ (inferred from
series analysis), an interesting question is
whether, for values of $\mu$ in the interval $-1 < \mu < 0$, the singularities
in $C$, $M$, and $\bar\chi$ at the inner endpoint of the left line segment are
described by exponents which are (i) independent of $\mu$ in this interval, and
(ii) equal to their values at $\mu=-1$.   In Table 3 we list our results for
$\mu=-0.5$ and $-0.2$ (the results for $\mu=-0.1$ were less precise).
The values for $\alpha_{lhe}'$ and $\gamma_{lhe}'$ from the dlog Pad\'e and
differential approximants agreed very well with each other.  The value of
$\beta_{lhe}$ was more difficult to extract from the series analysis (in
particular, the dlog Pad\'e approximants yield somewhat larger values than the
differential approximants); using only the (presumably more accurate) DA's, we
obtain the value given in Table 3.
To within the estimated uncertainties, these results are consistent with being
independent of $\mu$ for the two values of $\mu$ considered, and with
satisfying the equality $\alpha_{lhe}'+2\beta_{lhe}+\gamma_{lhe}=2$.  Regarding
question (ii), $\gamma_{lhe}'$ is slightly, but not decisively,
lower than the inferred
value at $\mu=-1$, $\gamma_e'=5/4$.  The exponent $\beta_{lhe}$ is
also consistent with the $\mu=-1$ value, $\beta_e=-1/8$, but the large
uncertainty in its determination precludes a sensitive test of this equality.
The values of $\alpha_{lhe}'$ are about $4\sigma$ above unity.  Assuming the
series analysis gives a reliable determination of this exponent, this means
that it is not equal to the $\mu=-1$ value, $\alpha_e'=1$.
Furthermore, it means that the density
of zeros does diverge at $u_{lhe}$, a conclusion already hinted at by the
behavior of the zeros calculated on finite lattices.

\begin{table}
\begin{center}
\begin{tabular}{|c|c|c|c|c|c|c|} \hline \hline & & & & & & \\
$\mu$ & $\alpha_{lhe}'$ & $\beta_{lhe}$ & $\gamma_{lhe}'$ &
        $\alpha_{rhe}'$ & $\beta_{rhe}$ & $\gamma_{rhe}'$ \\
& & & & & & \\
\hline \hline
$-0.5$ & 1.20(5)  & $-0.2(1)$ & 1.20(4) & 1.65(12) & $-0.35(20)$ &
1.00(15) \\ \hline
$-0.2$ & 1.18(5)  & $-0.2(1)$ & 1.19(5) & 1.4(1) & $-0.2(1)$ &
1.15(20)  \\ \hline
\hline
\end{tabular}
\end{center}
\caption{Values of the singular exponents $\alpha_{lhe}'$, $\beta_{lhe}$,
$\gamma_{lhe}'$, $\alpha_{rhe}'$, $\beta_{rhe}$, and $\gamma_{rhe}'$ at the
inner endpoints of the left- and right-hand line segments, $u_{lhe}$ and
$u_{rhe}$, respectively, as determined from analyses of low-temperature series
for $C$, $M$, and $\bar\chi$ using dlog Pad\'e and differential approximants.}
\label{table3}
\end{table}

   We have also extracted the critical exponents at the inner endpoint of
the right-hand line segment $u_{rhe}$
from series analysis.  Here there are two specific
questions to investigate.  First, are these exponents consistent with being
independent of $\mu$ in the range $-1 < \mu < 0$?  Second, recalling that, in
the limit $\mu \to -1$, the right line segment decreases in length and
eventually disappears at $u=1$ (as inferred from our finite-lattice
calculations, in agreement with our series analysis), one would like to see
whether the exponents on the inner endpoint of this right-hand line
segment, $u_{rhe}$, might be equal to those which we found for the singularity
at $u=1$ for $\mu=-1$, viz., $\alpha_1'=0$ (finite logarithmic
non-analyticity in $C$), $\beta_1=-1/4$ (both exact), and $\gamma_1'=5/2$
(inferred from series analysis) \cite{ih}.  Our results are shown in Table 3.
Just as was true of
the inner endpoint of the left line segment, it was more difficult to get an
accurate value of $\beta_{rhe}$, and again we have quoted a value weighted more
by the differential approximants, with a commensurately large uncertainty
assigned.
These values of critical exponents at the inner endpoint of the right-hand line
segment are crudely consistent, to within large uncertainties, with being
independent of $\mu$ over the indicated range of $\mu$. They are also in accord
with the hypothesis that
$\alpha_{rhe}'+2\beta_{rhe}+\gamma_{rhe}'=2$.
As $\mu \to -1$, we find that it is more difficult to get accurate values for
the singular exponents.  This is not surprising, since in this limit the right
line segment disappears.  Regarding the second question, although the
uncertainties in the critical exponents are large, our results indicate that,
for the range of $\mu$ considered, the exponents $\alpha_{rhe}'$ and
$\gamma_{rhe}'$ are not equal to the respective exponents at
$u=1$, $\mu=-1$.  We cannot draw a firm conclusion concerning whether over the
same range of $\mu$, $\beta_{rhe}$ is equal or unequal to $\beta_1$.  The
value(s) of $\alpha_{rhe}'$ is about 4-5 $\sigma$ above 1, which supports the
conclusion that the exact value of this exponent is, indeed, $> 1$, and hence
that the density $g$ of zeros diverges at $u_{rhe}$, as it does at
$u_{lhe}$.  This is again consistent with the observed behavior of the zeros
calculated on finite lattices.

\section{Comments on Other Lattices}

\subsection{Honeycomb Lattice}

    The complex-temperature phase diagram of the zero-field
Ising model on a honeycomb
lattice was determined in Ref. \cite{chitri} from exact results.  In the
complex $z$ plane the boundaries consist of the union of an arc of the unit
circle $z=e^{i\theta}$, for $\pi/3 \le \theta \le 5\pi/3$ with a closed curve
lying in the region $Re(z) \ge 0$ which intersects the circle at $z=\pm i$
and the positive real $z$ axis at $z_c$ and $1/z_c$, where
$z_c=2-\sqrt{3}$.  This phase diagram consists of three complex-temperature
phases: (i) FM, surrounding the origin, (ii) PM around $z=1$, and (iii) AFM,
the outermost phase, extending to complex infinity. The complex-temperature
singularities were determined exactly for the specific heat and magnetization,
and were studied using low-temperature series expansions for the
susceptibility \cite{chitri}.

   Here we determine the complex-temperature phase diagram for the honeycomb
lattice at $\mu=-1$.  One first uses the identity
$e^{i(\pi/2)\sigma_n} = i\sigma_n$
to relate the model to one with no magnetic field and the partition function
\beq
Z=i^{N_s}\sum_{\sigma_n}(\prod_n \sigma_n)
e^{\sum_{<nn'>}\sigma_n K_{nn'}\sigma_{n'}}
\label{zrelhc}
\eeq
(where we have allowed the possibility that the $K_j$ along the three different
lattice vectors are different).
One can next associate dimers with pairs of $\sigma$'s at neighboring sites and
then re-exponentiate these to express the partition function in terms of a
model with the associated coupling $K_j$ shifted by $i\pi/2$
\cite{merlini,lw}. A
complete dimer covering is provided by placing these dimers along one of the
three lattice vectors on the honeycomb lattice.  The situation is particularly
simple, for the following reason: given that the coordinaton number is odd
($q=3$), one can multiply by $1=\sigma^2$ at each vertex and thereby place the
dimers on every bond of the lattice, carry out the re-exponentiation, and thus
map the model to one with the couplings along each of the three lattice vectors
shifted as $K_j \to K_j+i\pi/2$, $j=1,2,3$.  Since the shift is the same for
each of these lattice directions, one can immediately specialize to the case of
isotropic couplings.  As a result, the reduced free energy $f$
for $h=i\pi/2$ is
given, up to an additive term $i\pi/4$, simply by that for $h=0$ with the
replacement $K \to K+i\pi/2$, or equivalently, $z \to -z$ or $v \to 1/v$. This
is also true of $C$, $M$, $M_{st}$, and $\bar\chi$.
Hence, we can directly take over our results for the zero-field case, and we
find the complex-temperature phase diagram shown in Fig. 7.  This consists of
the FM, and AFM phases together with a phase in which $M=M_{st}=0$ identically,
whence the label ZM (standing for ``zero uniform and staggered
magnetization'').  The magnetization satisfies
$M(z,h=i\pi/2)=M(-z,h=0)$, and similarly for the staggered magnetization.
The points in the ZM phase are in 1--1 correspondence, under
$z \to -z$, with the points in the PM phase of the zero--field model.  The
reason that we do not label it as PM is that the nonzero magnetic field
explicitly breaks the Z$_2$ symmetry, so that there is no PM phase in the
strict sense.  However, as discussed in Ref. \cite{ih}, this non-invariance is
manifested only in the non-invariance of the constant $i\pi/4$ in $f$, and
hence does not affect derivatives such as $C$, $M$, and $\bar\chi$.  We note
the contrast with the square lattice, where for $\mu=-1$ (and indeed, also for
all $\mu \ne 1$), there is no phase in which $M$ and $M_{st}$ both vanish;
instead of becoming the ZM phase, the PM phase disappears completely.
Our determination of complex-temperature singularities in
$\bar\chi$ in Ref. \cite{chitri} can also be taken over directly to the present
case, with just the change $z \to -z$.  The phase diagram in $v$ shows the
interesting feature that the FM and AFM phases are separated by a line
of singular points on the imaginary $v$ axis.  This is similar to what we found
for the square lattice (see Fig. 2(b) of Ref. \cite{ih})  (A difference
is that in the case of the square lattice, the FM and AFM phases
extend to complex infinity in the right and left parts of the $v$ plane, and
hence the line separating them is the complete imaginary $v$ axis, whereas for
the honeycomb lattice, the FM and AFM phases are bounded by an oval-like curve
(Fig. 7(b)) outside of which is the ZM phase, and the line segment separating
them is finite, extending from $-i\sqrt{3}$ to $i\sqrt{3}$.)

    As $h$ increases from 0, the rigorous argument that one can analytically
continue outward from $z=0$ past the former FM-PM critical point at $z_c$,
to the vicinity of the former PM-AFM critical point at $1/z_c$ implies,
as in the case of the square lattice, that the right-hand boundary of the FM
phase breaks open at $z=z_c$, forming two complex-conjugate arcs which retract
away from the real axis.  The former PM phase becomes part of the enlarged FM
phase, again as in the square lattice case.  The FM phase is completely
separated from the AFM phase, and the former second-order critical point at
$1/z_c$ which separated the PM and AFM phases for $h=0$ now becomes a
first-order phase boundary between the FM and AFM phases.  The general
arguments given above imply that as $\mu \to 0^+$ (and also for $\mu \to
0^-$), the FM phase expands to fill the entire $z$ plane.  Finally, for the
special value $\mu=-1$, the model again exhibits a phase with $M=M_{st}=0$, the
ZM phase.

   It may be noted that the simple relation under $z \to -z$ between the free
energy and other thermodynamic functions which holds for this lattice also
holds for heteropolygonal lattices with odd coordination number.  Hence the
complex-temperature phase diagrams for the $3 \cdot 12^2$ and $4 \cdot 8^2$
lattices with $h=i\pi/2$ are simply given by the diagrams which we worked out
for the zero-field case \cite{cmo} with the replacement $z \to -z$ (and PM
$\to$ ZM).

\subsection{Triangular Lattice}

    We recall that for the zero-field Ising model on the triangular lattice,
the locus of complex-temperature points where the free
energy is non-analytic consists of the union of the circle
\beq
u = -\frac{1}{3} + \frac{2}{3}e^{i\theta}
\label{ucircle}
\eeq
(with $0 \le \theta < 2\pi$) and the semi-infinite line segment
\beq
-\infty \le u \le -\frac{1}{3}
\label{usegment}
\eeq
The FM phase is located within the circle, with the PM phase lying outside and
extending to complex infinity.

   By methods similar to those we have discussed, we calculate the
complex-temperature phase diagram for $h=i\pi/2$ and show it in Fig. 8.
In the $u$ plane, the continuous
locus of points where the free energy is non-analytic
consists of the union of the circular arc
\beq
u = \frac{1}{2}(-1 + e^{i\theta}) \ , \quad  \theta_{ce} \le |\theta| \le \pi
\label{ucirclem1}
\eeq
where
\beq
\theta_{ce}=arctan\biggl (\frac{4\sqrt{2}}{7}\biggr ) \simeq 38.9^\circ
\label{thetace}
\eeq
corresponding to the endpoints $u_{ce}=e^{i\theta_{ce}}$ and $u_{ce}^*$, where
\beq
u_{ce}=\frac{-1 + 2^{3/2}i}{9}
\label{uce}
\eeq
and the semi-infinite line segment
\beq
 -\infty \le u \le -\frac{1}{2}
\label{usegmentm1}
\eeq
Just as there was no AFM phase for $h=0$, so also there is none for
$h=i\pi/2$. As $h$ increases from 0, the PM phase abruptly disappears.  By the
same rigorous arguments as were discussed before, the right-hand boundary of
the FM phase breaks open at $u=u_c=1/3$ and the circle changes into a circular
arc, the complex-conjugate endpoints of which retract away from the positive
real axis.  All points in the $u$ (equivalently, $z$ or $v$) plane (not lying
on the above continuous locus of points where $f$ is non-analytic)
are analytically connected to each other and all lie in the FM phase.
The corresponding locus of points in the $v$ plane consists of the union of a
circular arc $v=e^{i\phi}$ traced out for $\phi_{ce} < \phi \le \pi$, where
$\phi_{ce}=arctan(2^{3/2}/3)=43.3^\circ$ with two complex-conjugate arcs
intersecting this circle at $v=\pm i$.  Comparing the $h=0$ and $h=i\pi/2$
complex-temperature phase diagrams in the $u$
plane, one sees that the former circle of radius 2/3
centered at $u=-1/3$ is replaced by an arc of a circle of radius 1/2
centered at $u=-1/2$, and the semi-infinite line segment
extending leftward from $u=-1/3$ is replaced by a similar line segment
starting at $-1/2$.

\section{Conclusions}

     In this paper we have determined the general features of the
complex-temperature phase diagram for the square-lattice Ising model in a
nonzero external magnetic field $0 < h \le \infty$ \ ($0 \le \mu < 1$).  Our
methods included calculations of complex-temperature zeros of the partition
function for finite lattices and analysis of low-temperature series.
We have also performed a similar analysis for $-1 < \mu \le 0$.  The results
enable one to exhibit a continuous connection, via the variation of $\mu$
through real values from 1 to $-1$, of the two known exact solutions of this
model, viz., those of Onsager and of Lee and Yang.  We have studied the
exponents in the specific heat,
magnetization, and susceptibility at certain complex-temperature singularities
which are present in the interval $-1 < \mu < 1$.
Extending our earlier work, we have also determined the exact
complex-temperature phase diagrams of the model at $\mu=-1$ on the honeycomb
and triangular lattices and have discussed the relation between these and the
zero-field phase diagrams.
Our results give further information about the still-intriguing issue of the
Ising model in a magnetic field.

This research was supported in part by the NSF grant PHY-93-09888.

\vfill
\eject

\begin{center}
{\bf Figure Captions}
\end{center}

 Fig. 1. \ Complex-temperature zeros of $Z$, calculated for $h=0$
($\mu=1$) on a $7 \times 8$ lattice with helical boundary conditions, as
compared with the exact result, in the $u=e^{-4K}$ plane.

 Fig. 2. \ Complex-temperature zeros of $Z$, calculated for $h=i\pi/2$
($\mu=-1$) on a $7 \times 8$ lattice with helical boundary conditions, as
compared with the exact result, in the $u=e^{-4K}$ plane.

 Fig. 3. \ Zeros of $Z$ in the $u$ plane, for $h$ varying from 0 to 1.25 in
increments of 0.05.  Lattice size and boundary conditions are as in Fig. 1.

 Fig. 4. \ Zeros of $Z$ in the $u$ plane, for $h$= (a) 0.25 (b) 0.5 (c) 0.75
(d) 1.0.  Lattice size and boundary conditions are as in Fig. 1.

 Fig. 5. \ Zeros of $Z$ in the $u$ plane for $-1 < \mu < 0$.
Denoting $h=i\pi/2 + h_r$, $h_r$
varies from 0.25 to 1.25 in increments of 0.05.
Lattice size and boundary conditions are as in Fig. 1.

 Fig. 6. \ Zeros of $Z$ in the $u$ plane for $\mu$= (a) $-0.5$ (b) $-0.2$
(c) $-0.1$.  Lattice size and boundary conditions are as in Fig. 1.

 Fig. 7. \ Complex-temperature phase diagram for the Ising model on the
honeycomb lattice with $\mu=-1$ in the variable (a) $z$ (b) $v$.

 Fig. 8. \ Complex-temperature phase diagram for the Ising model on the
triangular lattice with $\mu=-1$ in the variable $u$.

\vfill
\eject


\begin{thebibliography}{99}

\bibitem{ons}{L. Onsager, Phys. Rev. {\bf 65}, 117 (1944).}

\bibitem{yang}{C. N. Yang, Phys. Rev. {\bf 85}, 808 (1952).}

\bibitem{fisher}{M. E. Fisher, {\it Lectures in Theoretical Physics}
(Univ. of Colorado Press, Boulder, 1965), vol. 7C, p. 1.}

\bibitem{kat}{S. Katsura, Prog. Theor. Phys. {\bf 38}, 1415 (1967).}

\bibitem{chisq}{V. Matveev and R. Shrock, J. Phys. A {\bf 28}, 1557 (1995).}

\bibitem{ly}{T. D. Lee and C. N. Yang, Phys. Rev. {\bf 87}, 410 (1952).}

\bibitem{yl}{C. N. Yang and T. D. Lee, Phys. Rev.  {\bf 87}, 404 (1952).}

\bibitem{lp}{J. L. Lebowitz and O. Penrose, Commun. Math. Phys. {\bf 11}, 99
(1968).}

\bibitem{bdn}{H. W. Bl\"ote and M. den Nijs, Phys. Rev. {\bf B37},
1766 (1988).}

\bibitem{abe}{R. Abe, Prog. Theor. Phys. {\bf 38}, 322 (1967).}

\bibitem{oksk}{S. Ono, Y. Karaki, M. Suzuki, and C. Kawabata,
J. Phys. Soc. Jpn. {\bf 25}, 54 (1968)}

\bibitem{g69}{C. J. Thompson, A. J. Guttmann, and B. W. Ninham, J. Phys. C
{\bf 2}, 1889 (1969); A. J. Guttmann {\it ibid}, 1900 (1969).}

\bibitem{dg}{C. Domb and A. J. Guttmann, J. Phys. C {\bf 3}, 1652 (1970).}

\bibitem{ak}{Y. Abe and S. Katsura, Prog. Theor. Phys. {\bf 43}, 1402 (1970).}

\bibitem{ms}{G. Marchesini and R. Shrock, Nucl. Phys. {\bf B318}, 541 (1989).}

\bibitem{egj}{I. G. Enting, A. J. Guttmann, and I. Jensen, J. Phys. A
{\bf 27}, 6963 (1994).}

\bibitem{merlini}{D. Merlini, Lett. Nuovo Cim. {\bf 9}, 100 (1974).}

\bibitem{lw}{K. Y. Lin and F. Y. Wu, Int. J. Mod. Phys. B {\bf 4}, 471 (1988).}

\bibitem{ih}{V. Matveev and R. Shrock, J. Phys. A, in press
(hep-lat/9412105).}

\bibitem{ipz}{C. Itzykson, R. Pearson, and J. B. Zuber, Nucl. Phys.
{\bf B220} 415 (1983).}

\bibitem{fsm}{M. E. Fisher, Proc. Roy. Soc. A {\bf 254}, 66 (1960); {\bf 256},
502 (1966).}

\bibitem{aniso}{W. van Saarloos and D. Kurtze, J. Phys. A {\bf 17}, 1301
(1984); J. Stephenson and R. Couzens, Physica {\bf 129A}, 201 (1984);
D. Wood, J. Phys. A {\bf 18} L481 (1985); J. Stephenson and J. van Aalst,
Physica {\bf 136A}, 160 (1986).}

\bibitem{cmo}{V. Matveev and R. Shrock, J. Phys. A, in press
(hep-lat/9503005).}

\bibitem{chitri}{V. Matveev and R. Shrock, J. Phys. A, in press
(hep-lat/9411023, 9412076).}

\bibitem{tm}{K. Binder, Physica {\bf 62}, 508 (1972); T. de Neef and
I. G. Enting, J. Phys. A {\bf 10}, 801 (1977); I. G. Enting,
Aust. J. Phys. {\bf 31}, 515 (1978); G. Bhanot, J. Stat. Phys.
{\bf 60}, 55 (1990).}

\bibitem{tlow1}{M. F. Sykes, D. S. Gaunt, J. L. Martin, S. R. Mattingly, and
J. W. Essam, J. Math. Phys. A {\bf 14}, 1071 (1973).}

\bibitem{tlow2}{M. F. Sykes, M. G. Watts, and D. S. Gaunt, J. Math. Phys. A
{\bf 8}, 1448 (1975).}

\bibitem{be}{R. J. Baxter and I. G. Enting, J. Stat. Phys. {\bf 21},
103 (1979).}

\bibitem{tonyg}{A. J. Guttmann, in {\it Phase Transitions and Critical
Phenomena}, Domb, C. and Lebowitz, J., eds. (Academic Press, New York, 1989)
vol. 13.}


\end{thebibliography}
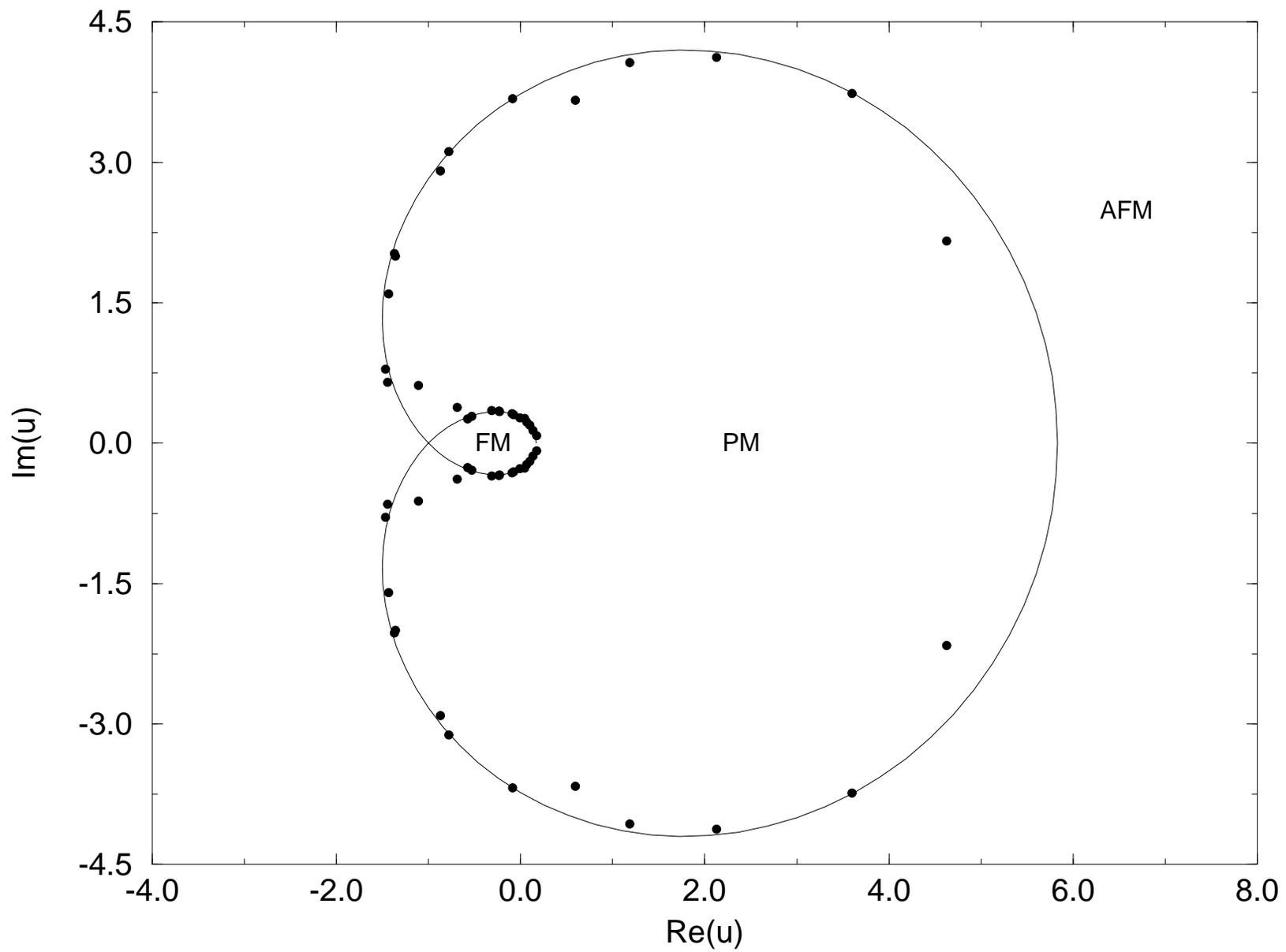
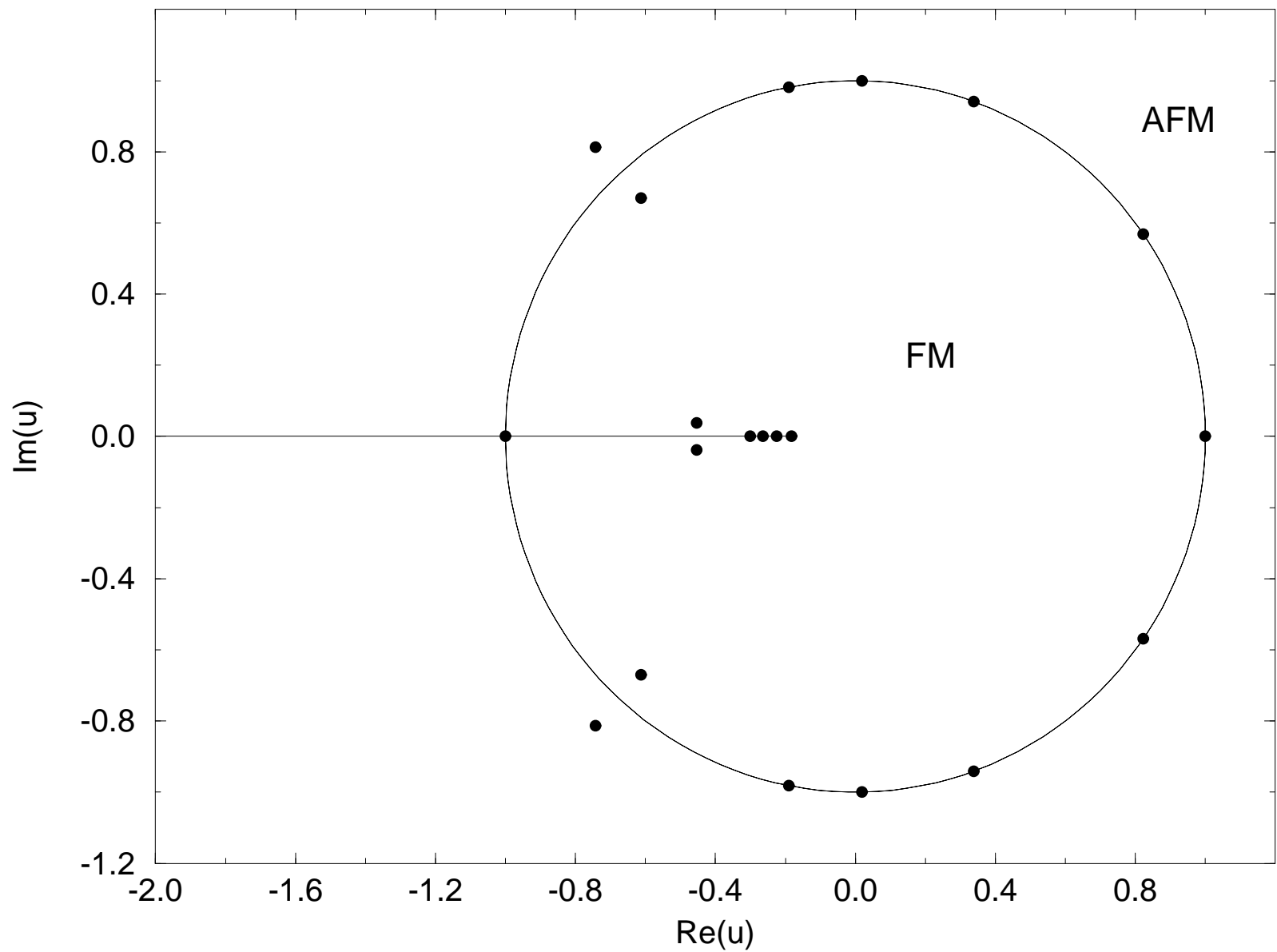
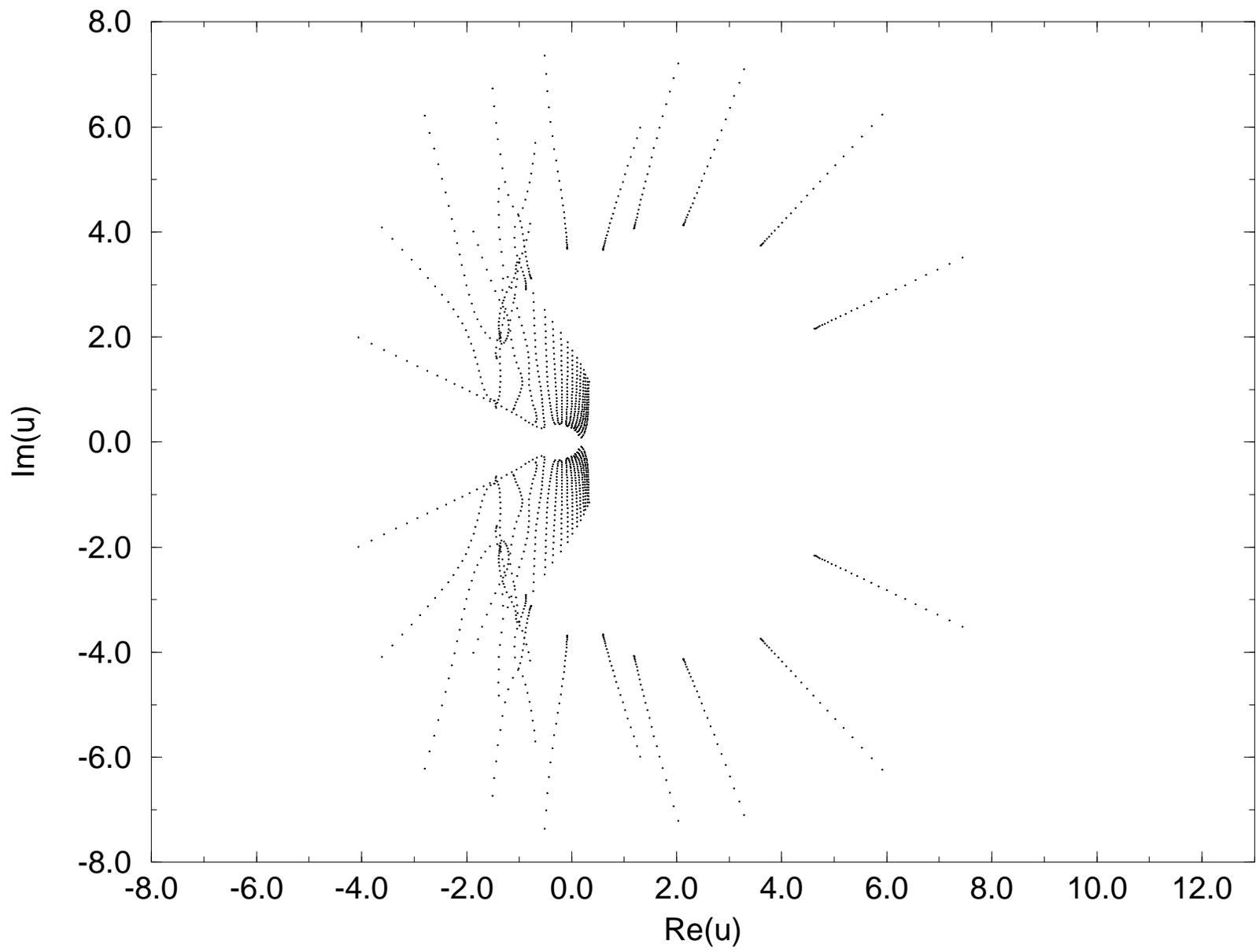
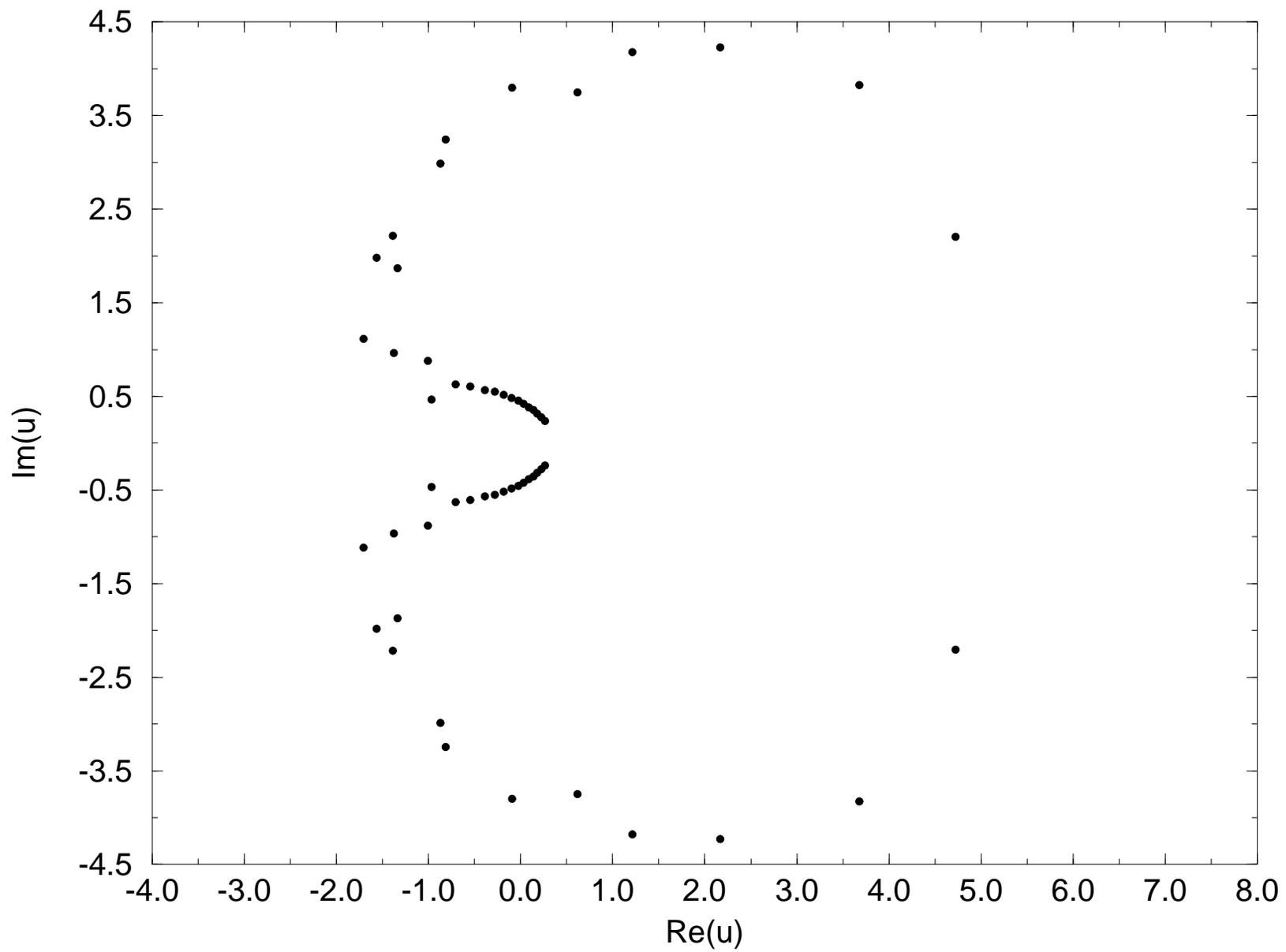
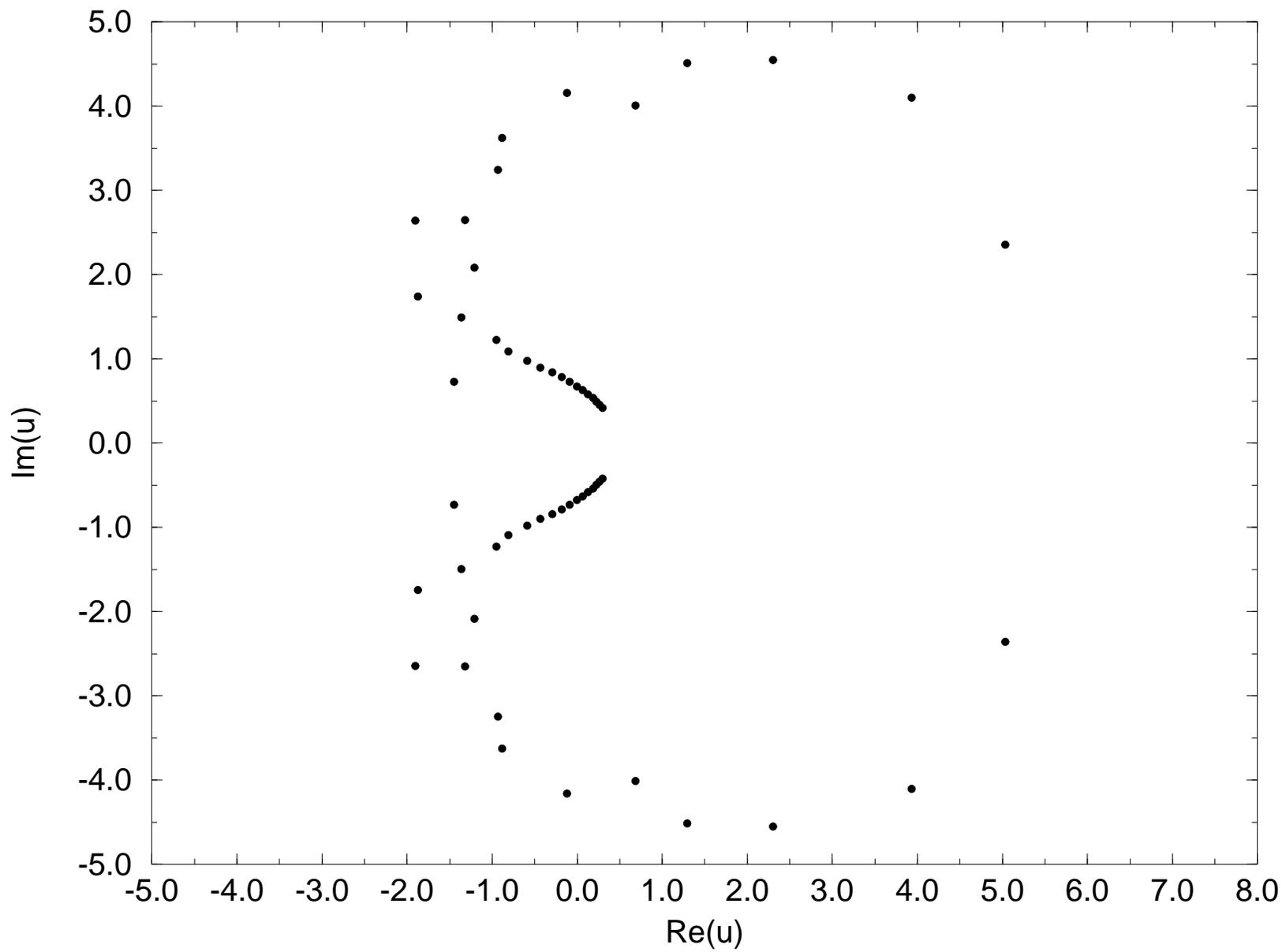
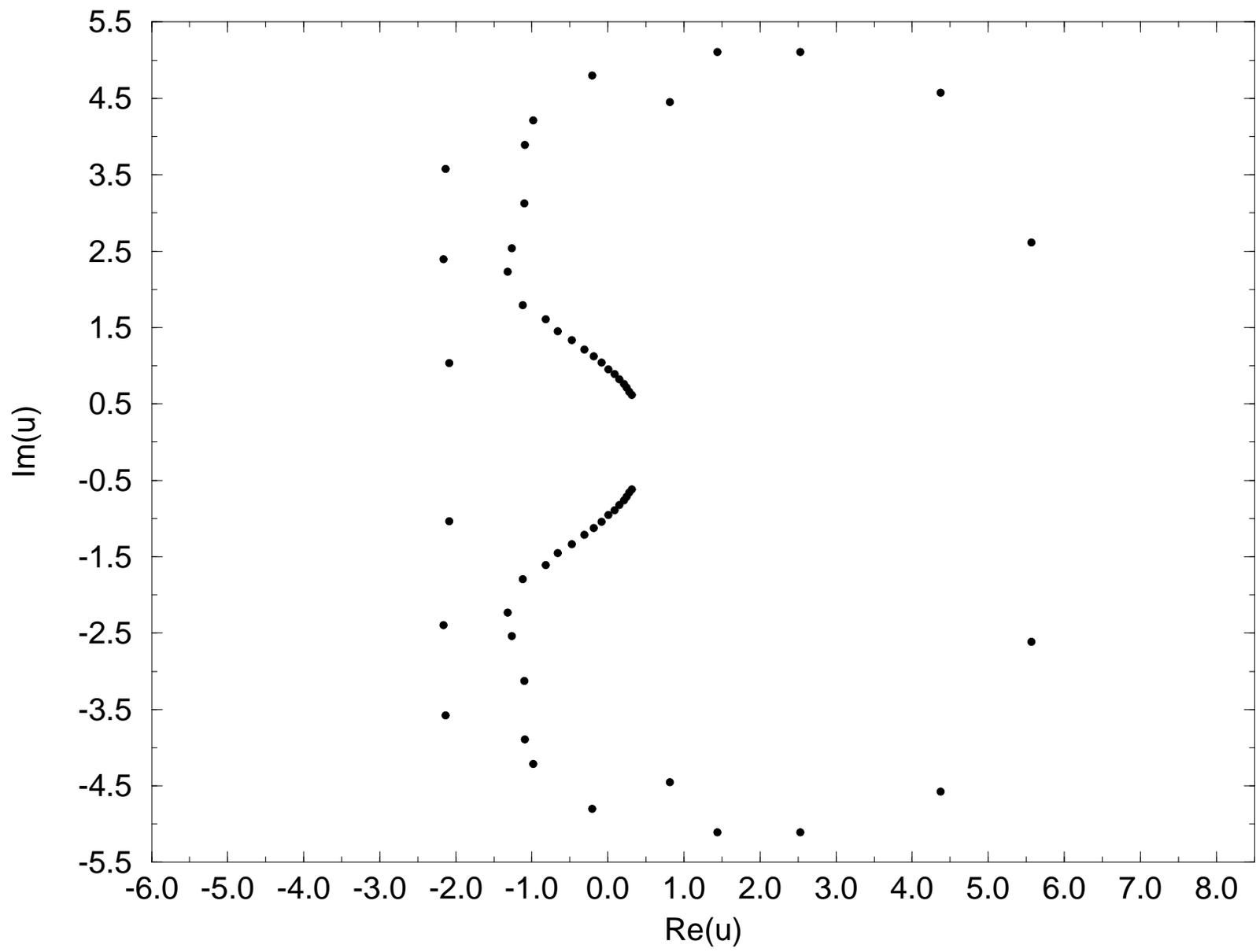
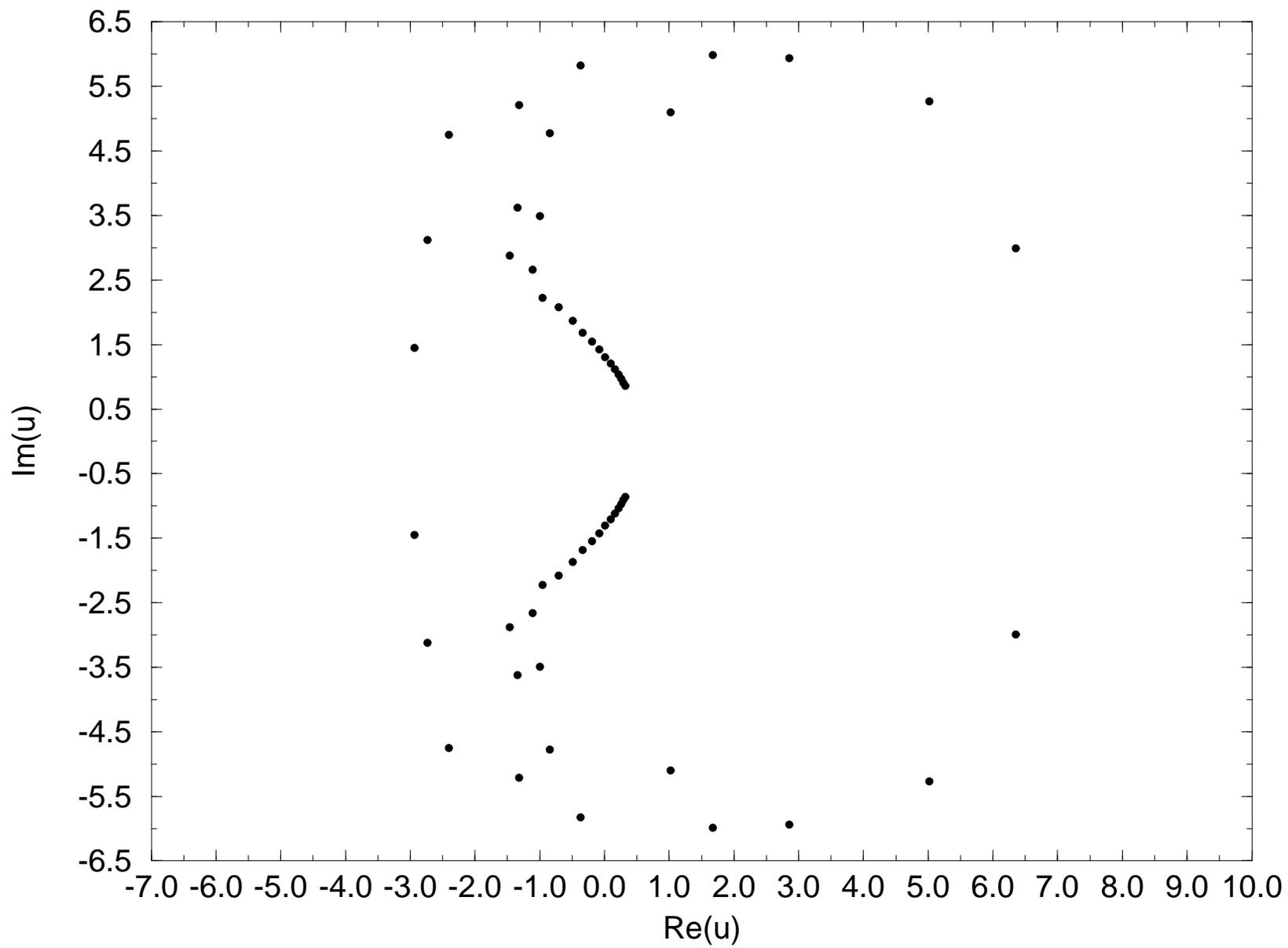
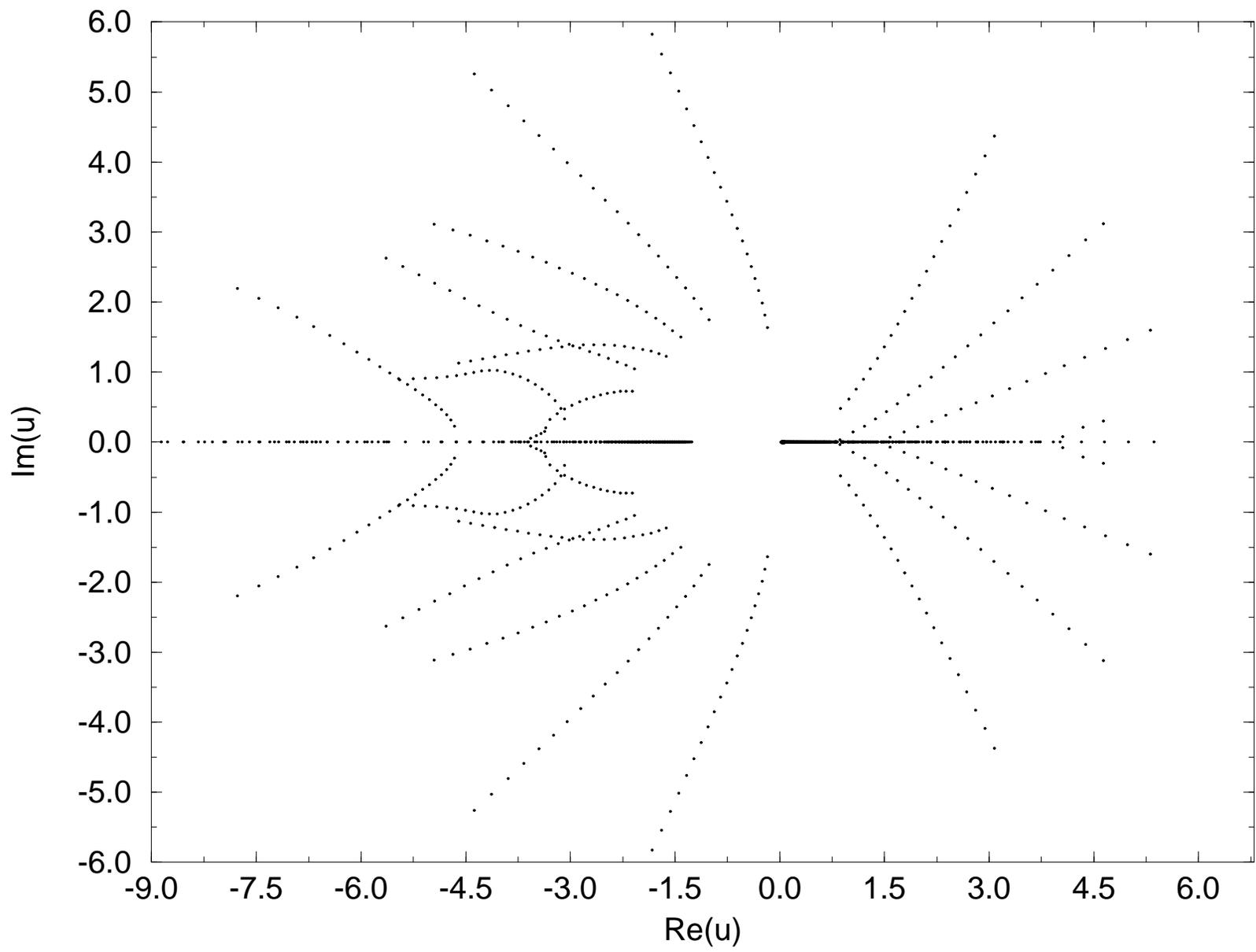
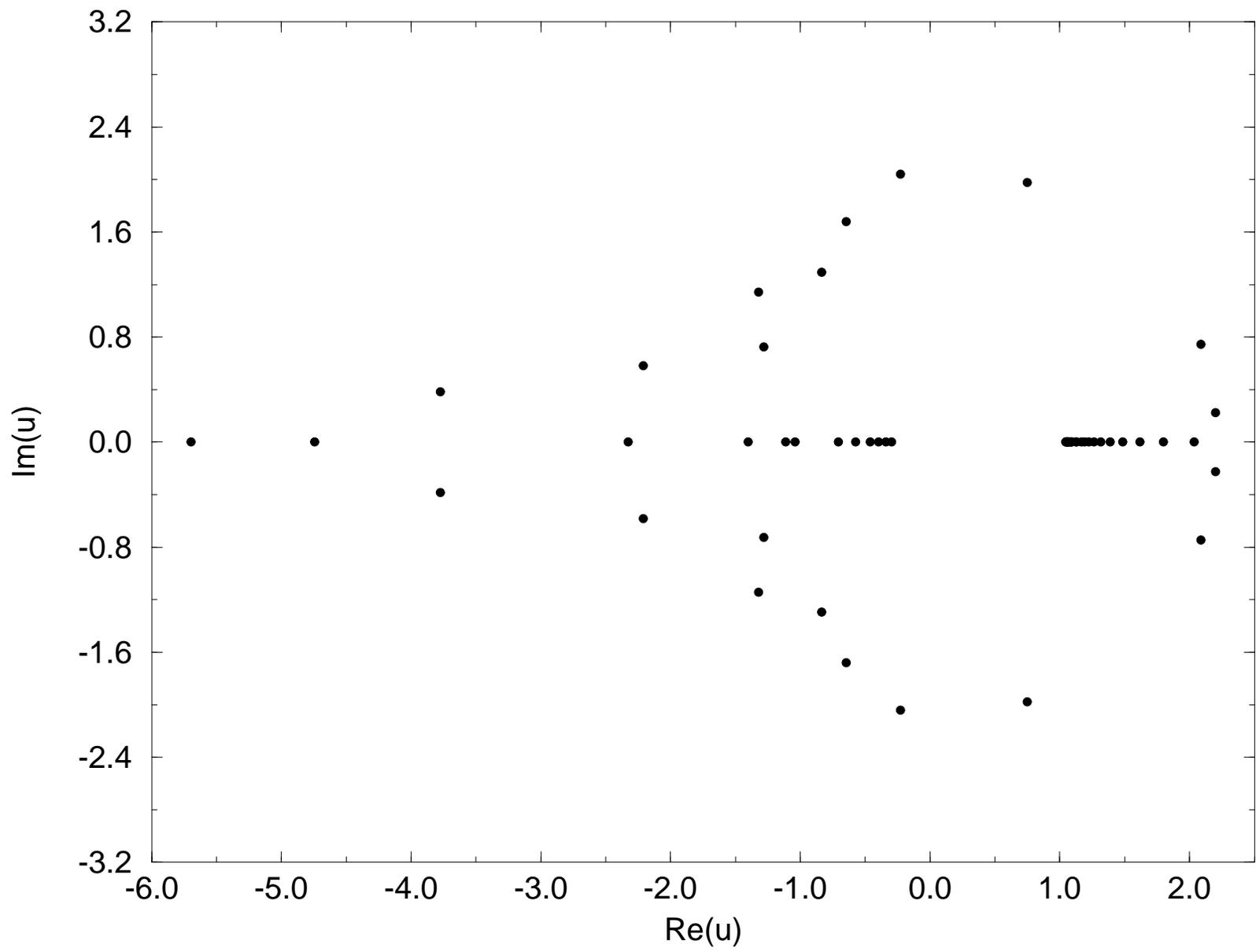
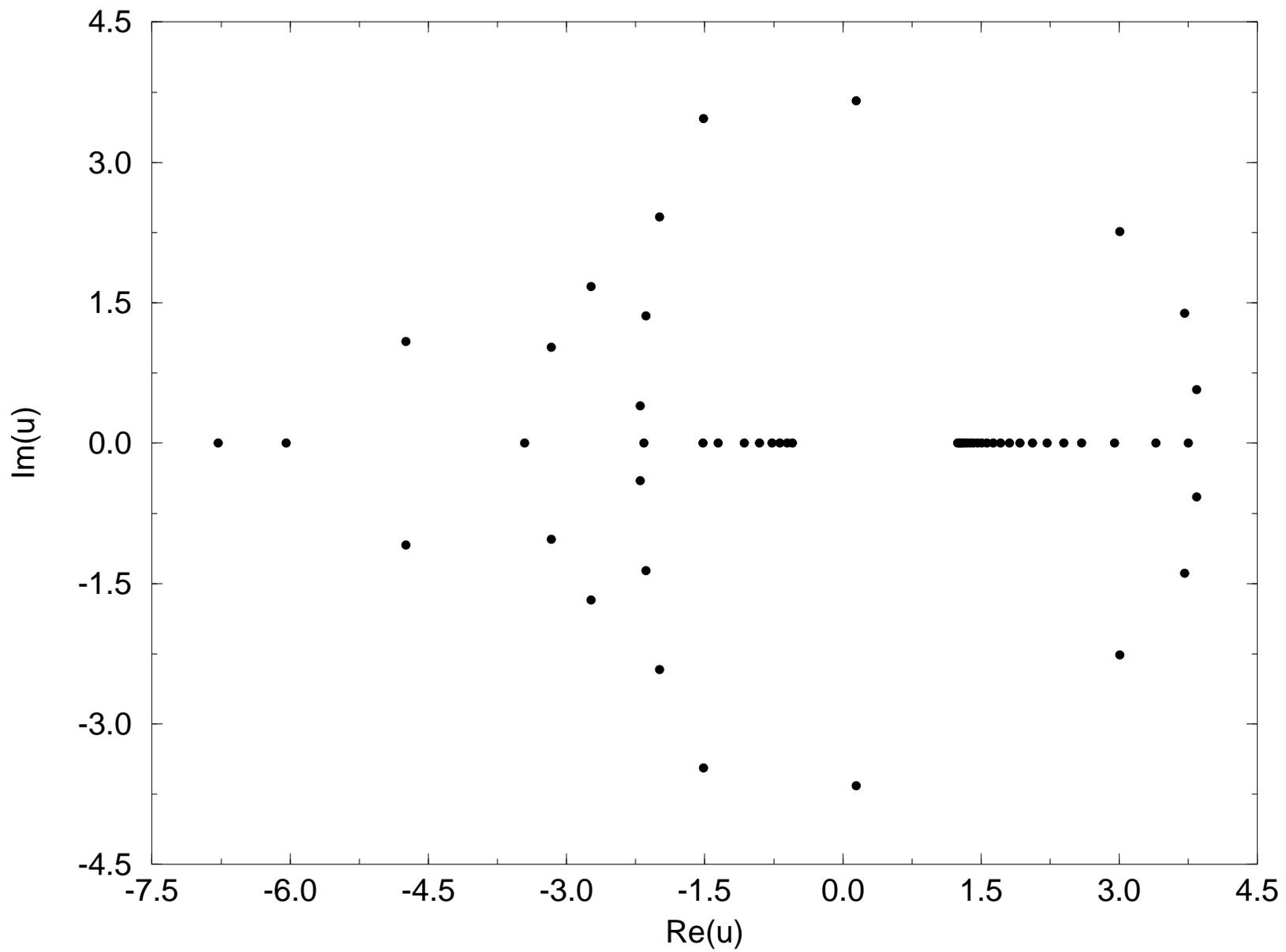
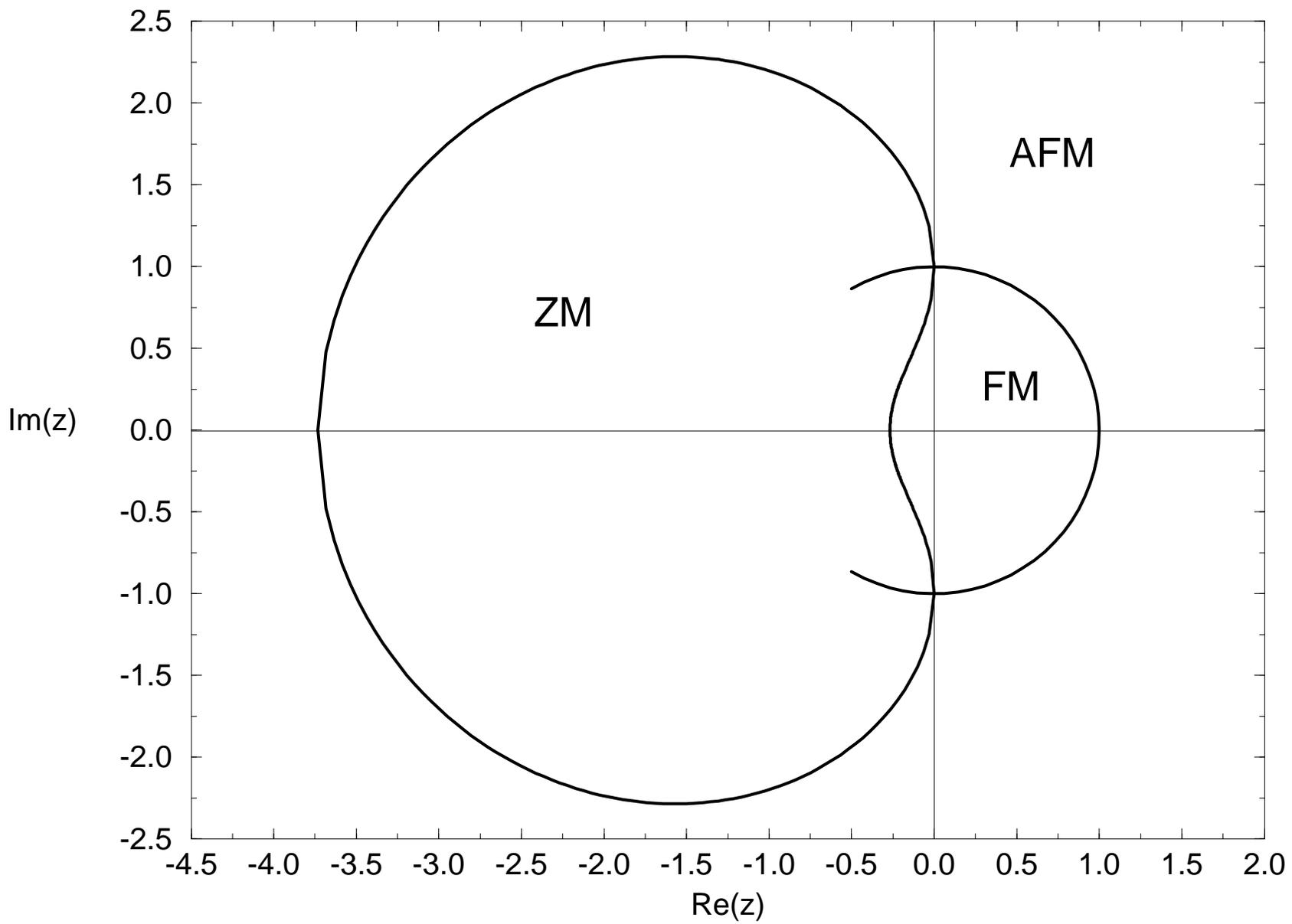
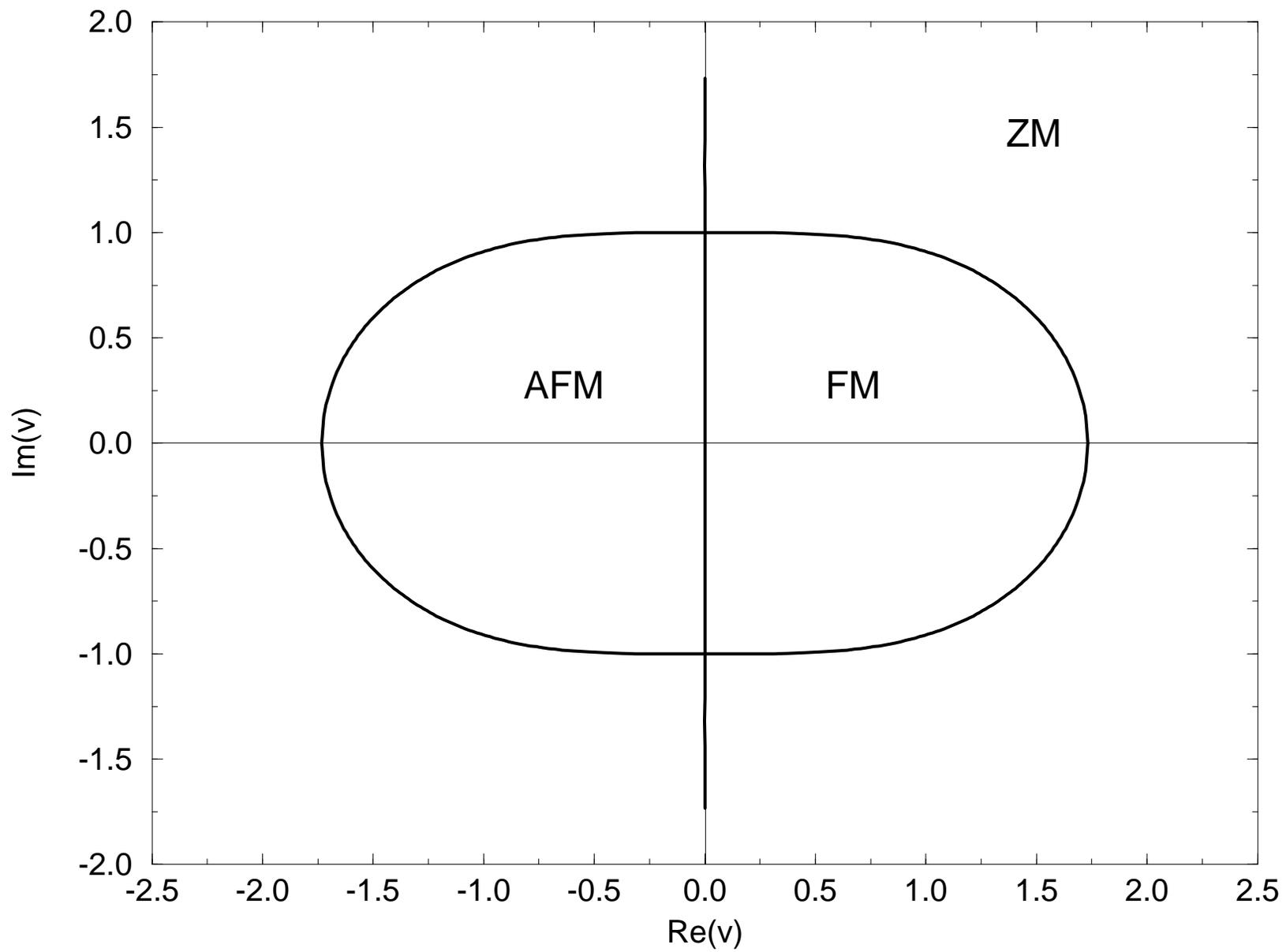
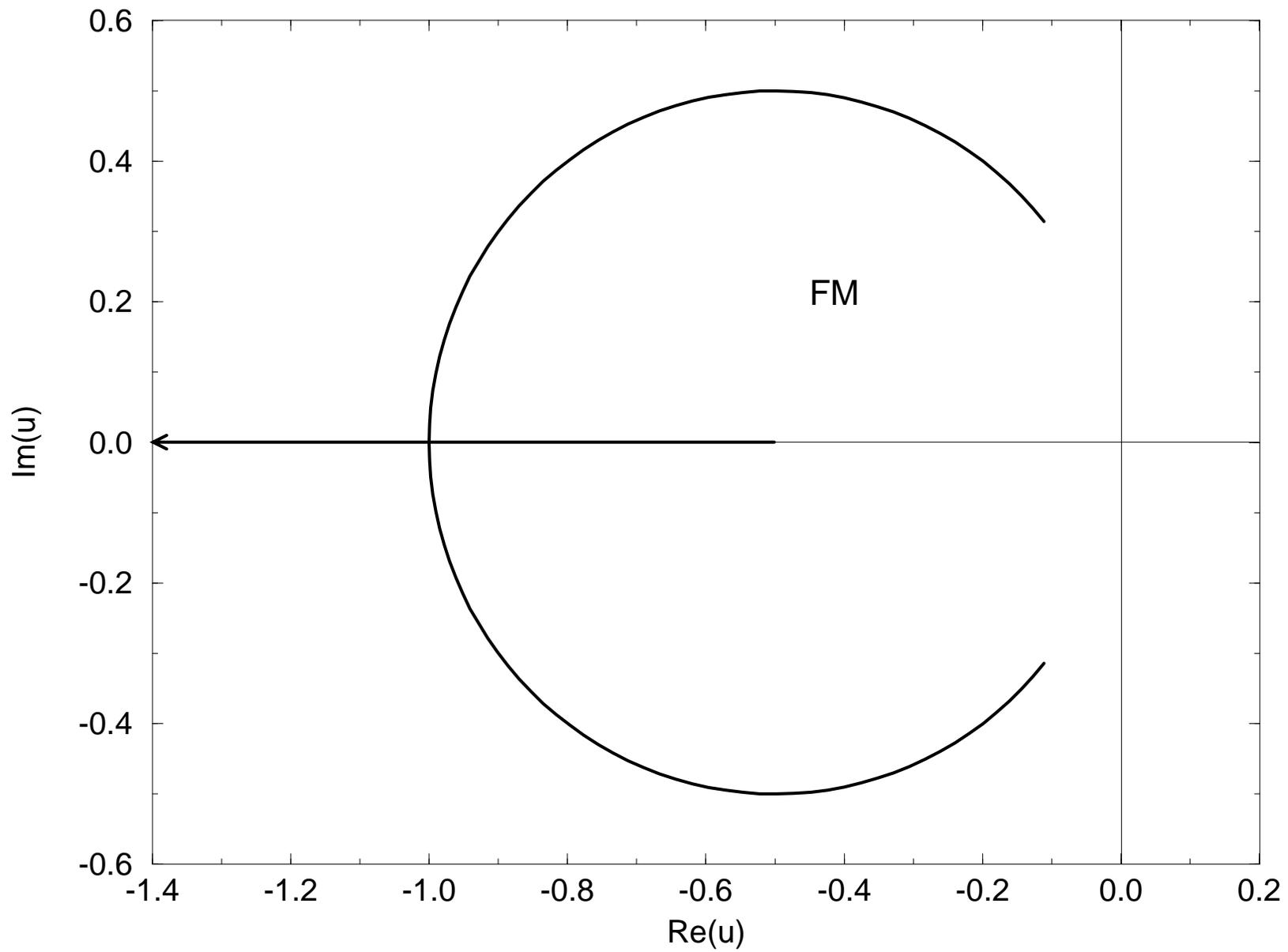
\end{document}